\documentclass[pre,aps,10pt,superscriptaddress,twocolumn,floatfix]{revtex4-1}

\usepackage[nice]{nicefrac}
\usepackage{graphicx,color}
\usepackage{amsmath,amssymb,bm,bbm}
\usepackage{amsmath}
\usepackage{braket}
\usepackage{float}
\usepackage{placeins}

\usepackage[plainpages=false,pdfpagelabels,colorlinks=true,linkcolor=red,urlcolor=blue,citecolor=blue,pdftitle={},pdfauthor={},pdfdisplaydoctitle=true,pdfduplex=DuplexFlipLongEdge]{hyperref}

\definecolor{darkred}{rgb}{0.90,0.2,0.2}
\definecolor{darkgreen}{rgb}{0,0.60,.2}
\definecolor{darkblue}{rgb}{0.1,0.3,1}
\definecolor{grey}{cmyk}{0,0,0,0.25}
\definecolor{orange}{cmyk}{0,0.6,0.8,0}

\begin{document}

\title{Single-quasiparticle eigenstate thermalization}

\author{Piotr Tokarczyk}
\affiliation{Institute of Theoretical Physics, Wroclaw University of Science and Technology, 50-370 Wroc{\l}aw, Poland}
\author{Lev Vidmar}
\affiliation{Department of Theoretical Physics, J. Stefan Institute, SI-1000 Ljubljana, Slovenia}
\affiliation{Department of Physics, Faculty of Mathematics and Physics, University of Ljubljana, SI-1000 Ljubljana, Slovenia\looseness=-1}
\author{Patrycja  \L yd\.{z}ba}
\affiliation{Institute of Theoretical Physics, Wroclaw University of Science and Technology, 50-370 Wroc{\l}aw, Poland}

\begin{abstract}
Quadratic Hamiltonians that exhibit single-particle quantum chaos are called quantum-chaotic quadratic Hamiltonians.
One of their hallmarks is single-particle eigenstate thermalization introduced in \href{https://doi.org/10.1103/PhysRevB.104.214203}{Phys.~Rev.~B~{\bf 104},~214203~(2021)}, which describes statistical properties of matrix elements of observables in single-particle eigenstates.
However, the latter has been studied only in quantum-chaotic quadratic Hamiltonians that obey the U(1) symmetry.
Here, we focus on quantum-chaotic quadratic Hamiltonians that break the U(1) symmetry and, hence, their "single-particle" eigenstates are actually single-quasiparticle excitations introduced on the top of a many-body state.
We study their wave functions and matrix elements of one-body observables, for which we introduce the notion of {\it single-quasiparticle} eigenstate thermalization.
Focusing on spinless fermion Hamiltonians in three dimensions with local hopping, pairing and on-site disorder, we also study the properties of disorder-induced near zero modes, which give rise to a sharp peak in the density of states at zero energy.
Finally, we numerically show equilibration of observables in many-body eigenstates after a quantum quench. We analytically prove that it is a consequence of single-quasiparticle eigenstate thermalization, in analogy to the U(1) symmetric case from \href{https://journals.aps.org/prl/abstract/10.1103/PhysRevLett.131.060401}{Phys. Rev. Lett. {\bf 131}, 060401 (2023)}.
\end{abstract}
\maketitle

\section{Introduction}

The emergence of quantum chaos in single-particle systems is nowadays a well established concept.
Inspired by some early contributions such as Berry's conjecture for the structure of chaotic wavefunctions~\cite{berry_77} and the quantum chaos conjecture for the statistics of Hamiltonian eigenvalues~\cite{casati_valzgris_80, bohigas_giannoni_84}, signatures of chaos were detected in various physical systems~\cite{stoeckmann_99, haake_gnutzmann_18, dalessio_kafri_16, borgonovi_izrailev_16}.
In the context of quadratic lattice Hamiltonians, such as the Anderson model, a comparison of single-particle spectral statistics with random matrix theory predictions served as one of the key tools for a detection of chaos and its breakdown at the localization transition~\cite{shklovskii_shapiro_93, hofstetter_schreiber_93, tarquini_biroli_17, suntajs_prosen_21, sierant_lewenstein_22, garciamata_martin_22, suntajs_prosen_23}.

Recent works have established a broad characterization of single-particle chaos in quadratic systems beyond single-particle spectral statistics~\cite{lydzba_rigol_21, lydzba_zhang_21, ulcakar_vidmar_22}.
For example, closed-form expressions for the average many-body eigenstate entanglement entropies of these systems were introduced~\cite{Lydzba_2020, lydzba_rigol_21} (see also~\cite{bianchi_hackal_21, bianchi_hackl_22}), and the validity of eigenstate thermalization in single-particle eigenstates was established~\cite{lydzba_zhang_21, ulcakar_vidmar_22}.
Quadratic Hamiltonians that comply with these properties were named {\it quantum-chaotic quadratic} Hamiltonians~\cite{lydzba_rigol_21}.

We note that the single-particle eigenstate thermalization refers to the statistical properties of matrix elements of observables in single-particle eigenstates, and does not imply thermalization of quadratic Hamiltonians.
Since generic single-particle states have broad distributions of single-particle energies and there is no clear definition of the microcanonical ensemble for a single-particle system, in this work we do not attempt to make any relation of the single-particle eigenstate thermalization with the thermalization of a single-particle system.

Perhaps the most important aspect of single-particle eigenstate thermalization ansatz is its relevance for quantum dynamics of many-body states. 
We note that this ansatz does not imply thermalization of quadratic systems with a non-vanishing particle density.
However, it has been shown that its validity guarantees equilibration of observables after quantum quenches~\cite{lydzba_mierzejewski_23}.
Equilibration of observables in quadratic systems~\cite{Gramsch_2012, Ziraldo_2012, Ziraldo_2013, He_2013, wright_rigol_14, Gluza_2016, Gluza_2019, murthy19} represents a necessary condition to establish the agreement between their long time averages and the generalized Gibbs ensemble predictions~\cite{gge, vidmar16}.

It should be emphasized that the validity of single-particle eigenstate thermalization has so far only be established in quadratic systems with the particle number conservation, i.e., with the U(1) symmetry.
The single-particle description of general (i.e., particle number nonconserving) quadratic Hamiltonians is commonly described in terms of quasiparticles or quasiholes created on the top of some vacuum state~\cite{Sato_2017,Kinnunen_2018}.
However, this vacuum state typically consists of states with many bare fermions, and hence it represents a many-body state.
An open question is then to establish the analog of single-particle eigenstate thermalization in general quadratic Hamiltonians.

In this paper, we achieve this goal and introduce the notion of {\it single-quasiparticle} eigenstate thermalization.
We study a quantum-chaotic quadratic Hamiltonian for spinless fermions in a cubic lattice that includes local hopping, pairing and weak on-site disorder.
We show that the single-quasiparticle (and single-quasihole) wavefunctions exhibit features of quantum chaos, and are hence consistent with Berry's conjecture~\cite{berry_77}.
A special attention is devoted to the analysis of disorder-induced near zero modes, which can be detected by a sharp peak in the density of states.
However, the relative number of these states vanishes in the thermodynamic limit.

We then show that observables in the overwhelming majority of single-quasiparticle eigenstates (away from zero modes) exhibit key properties of eigenstate thermalization, dubbed single-quasiparticle eigenstate thermalization.
In the latter, fluctuations of the observable matrix elements decay as a square root of the number of lattice sites $V$.
This should be contrasted to the eigenstate thermalization hypothesis (ETH) emerging in many-body eigenstates of interacting systems~\cite{deutsch_91, srednicki_94, rigol_dunjko_08, dalessio_kafri_16, mori_ikeda_18, deutsch_18}, which describes an exponential in $V$ decay of the observable matrix elements fluctuations~\cite{rigol_dunjko_08, steinigeweg2013, Beugeling2014, Kim_strong2014, Mondaini2016, yoshizawa_iyoda_18, jansen_stolpp_19, leblond_mallayya_19, mierzejewski_vidmar_20, richter_dymarsky_20, schoenle_jansen_21, sugimoto_hamazaki_21, noh_21, brenes_pappalardi_21}.
A direct implication of single-quasiparticle eigenstate thermalization is equilibration of observables after quantum quenches, which is shown analytically and tested numerically. 

The paper is organized as follows.
In Sec.~\ref{sec:model} we introduce the system under investigation and discuss its general properties.
In Sec.~\ref{sec:wavefunction} we study the properties of single-quasiparticle and single-quasihole wavefunctions, with a special emphasis on the disorder-induced near zero modes.
We then study in Sec.~\ref{sec:sqETH} the statistical properties of diagonal and off-diagonal matrix elements of observables in these states, thereby establishing the key properties of single-quasiparticle eigenstate thermalization.
Finally, in Sec.~\ref{sec:quench} we analytically and numerically study the time evolution of observables after quantum quenches, and show that single-quasiparticle eigenstate thermalization guarantees equilibration of observables.
We conclude in Sec.~\ref{sec:conclusions}.

\section{Model and Set-up} \label{sec:model}

\subsection{General considerations} \label{sec:general}

We are interested in a general form of a quadratic Hamiltonian for spinless fermions on a lattice with $V$ sites,
\begin{equation}
\label{eqH}
\hat{H}=\sum_{i,j=1}^V \left( h_{ij} \hat{c}_i^\dagger \hat{c}_j + \varDelta_{ij}\hat{c}_i^\dagger \hat{c}_j^\dagger + \varDelta_{ij}^* \hat{c}_j \hat{c}_i \right) \;,
\end{equation}
where $\hat{c}_{i}^\dagger$ ($\hat{c}_{i}$) creates (annihilates) a spinless fermion at site $i$.
The first term in Eq.~(\ref{eqH}) describes the hopping between sites $i$ and $j$ (when $i\neq j$) and the on-site potential (when $i=j$), while the other terms represent the pairing that breaks the particle number conservation.
The coefficients $h_{ij} = h_{ji}^*$ form a hermitian $V \times V$ matrix $h$, while the coefficients $\varDelta_{ij}=-\varDelta_{ji}$ form an antisymmetric $V \times V$ matrix $\varDelta$.

For the sake of completeness, in this section we describe some general properties of the Hamiltonian $\hat{H}$ from Eq.~(\ref{eqH}), while in Sec.~\ref{sec:ourmodel} we introduce the specific model under investigation.
Diagonalization of $\hat{H}$ is usually carried out by adopting the Nambu representation,
\begin{equation}
\label{eqH2}
    \hat{H} = \hat{C}^\dagger
    \begin{bmatrix}
    \frac{1}{2}{h} & \varDelta\\
    -\varDelta^{*} & -\frac{1}{2}h^{*}
    \end{bmatrix}\hat{C} + \sum_{i=1}^V \frac{h_{ii}}{2}
    =\hat{C}^\dagger \mathcal{H} \hat{C} + \sum_{i=1}^V \frac{h_{ii}}{2},
\end{equation}
where $\hat{C}=\left[\hat{c}_{1}\;...\;\hat{c}_{V}\;\hat{c}_{1}^\dagger\;...\;\hat{c}_{V}^\dagger\right]^\text{T}$ is a $2V\times 1$ vector.
The $2V \times 2V$ matrix $\mathcal{H}$ is said to be the matrix representation of the Hamiltonian $\hat{H}$ in the Bogoliubov-de Gennes basis.
For a convenience, in what follows we omit the constant term in Eq.~(\ref{eqH2}), i.e., we set $\sum_i \frac{h_{ii}}{2}=0$.

We introduce a unitary transformation to diagonalize $\mathcal{H}$ as $\hat{C}^\dagger \mathcal{H} \hat{C} = \hat{F}^\dagger \mathcal{D} \hat{F}$, where $\mathcal{D} = U^\dagger \mathcal{H} U$ is a $2V \times 2V$ diagonal matrix and $\hat{F} = U^\dagger \hat{C}$ contains the new fermionic annihilation and creation operators, where $\hat{F}=\left[\hat{f}_{1}\;...\;\hat{f}_{V}\;\hat{f}_{1}^\dagger\;...\;\hat{f}_{V}^\dagger\right]^\text{T}$ is a $2V\times 1$ vector.
The $2V \times 2V$ unitary matrix $U$ can be expressed as
\begin{equation} \label{eq:def_U}
    U=
    \begin{bmatrix}
    u & v^*\\
    v & u^*
    \end{bmatrix},
\end{equation}
where $u$ and $v$ are $V \times V$ matrices with matrix elements $u_{i\alpha}$ and $v_{i\alpha}$.
The form of $U$ from Eq.~(\ref{eq:def_U}) gives rise to a special property of the diagonal matrix $\mathcal{D}$, i.e., the diagonal matrix elements are $\{\epsilon_\alpha\}$ for $\alpha=1,...,V$, and the remaining elements are $\{-\epsilon_\alpha\}$.

The new fermionic creation and annihilation operators can be expressed as~\cite{Bogoliubov_1958,Fetter_2003,Kinnunen_2018}
\begin{equation} \label{eq:def_falpha}
\hat{f}_\alpha^\dagger = \sum_{i=1}^{V}(u_{i\alpha}\hat{c}_{i}^\dagger + v_{i\alpha} \hat{c}_{i}) \;,\;\;
\hat{f}_\alpha = \sum_{i=1}^{V}(u_{i\alpha}^{*}\hat{c}_{i}+v_{i\alpha}^{*}\hat{c}_{i}^\dagger)\;,
\end{equation}
while the annihilation operators of bare fermions in the site-occupation basis can be expressed using the inverse transformation as $\hat{c}_i = \sum_{\alpha=1}^{V}(u_{i\alpha}\hat{f}_{\alpha}+v_{i\alpha}^{*}\hat{f}_{\alpha}^\dagger)$. 
This brings the Hamiltonian $\hat{H}$ to the diagonal form, 
\begin{equation} \label{def_H_diag}
    \hat{H} =\sum_{\alpha=1}^{V}\left(\epsilon_\alpha \hat{f}_{\alpha}^\dagger\hat{f}_{\alpha}-\epsilon_\alpha \hat{f}_{\alpha}\hat{f}_{\alpha}^\dagger\right) 
    = \sum_{\alpha=1}^{V}2\epsilon_\alpha \hat{f}_{\alpha}^\dagger\hat{f}_{\alpha}-\sum_{\alpha=1}^{V}\epsilon_\alpha\;.
\end{equation}
One often refers to $\hat{f}_{\alpha}^\dagger$ and $\hat{f}_{\alpha}$ as the creation operators of the so-called Bogoliubov quasiparticles and quasiholes with energy $\epsilon_{\alpha}$ and $-\epsilon_{\alpha}$, respectively.
Since both $\hat{f}_{\alpha}^\dagger$ and $\hat{f}_{\alpha}$ are linear combinations of $\hat{c}_{i}^\dagger$ and $\hat{c}_{i}$, both quasiparticle and quasihole excitations are superpositions of the bare particle and hole excitations.

In principle one has no {\it a priori} knowledge of the sign of $\epsilon_\alpha$.
However, one notes that the columns $\alpha$ and $\alpha+V$ (assuming $\alpha \leq V$) of the unitary matrix $U$ in Eq.~(\ref{eq:def_U}) are related, which is a consequence of the relationship between $\hat F_\alpha^\dagger = \hat f_\alpha^\dagger$ and $\hat F_{\alpha+V}^\dagger = \hat f_\alpha$, see Eq.~(\ref{eq:def_falpha}).
In the matrix representation, the eigenstates of ${\cal H}$ can be expressed as
$\psi_\alpha = [u_{1\alpha}\, ...\, u_{V\alpha} \, v_{1\alpha}\, ...\, v_{V\alpha}]^{\rm T}$ and 
$\psi_{\alpha+V} = [v_{1\alpha}\, ...\, v_{V\alpha} \, u_{1\alpha}\, ...\, u_{V\alpha}]^\dagger$, satisfying
${\cal H} \psi_\alpha = \epsilon_\alpha$ and ${\cal H} \psi_{\alpha+V} = -\epsilon_\alpha \psi_{\alpha+V}$.
It is then convenient to assume $\epsilon_\alpha \ge 0$, which associates quasiparticles as objects with non-negative energies and quasiholes as objects with non-positive energies.

As a side remark, which is not essential for understanding of our new results, we note that in the case when quasiparticles are associated with energies $\epsilon_\alpha \ge 0$, one can introduce the ground state $|\Psi_0\rangle$ as the eigenstate with no quasiparticle excitations, with energy $E_0=-\sum_{\alpha=1}^{V}\epsilon_\alpha$.
Since $\hat{f}_{\alpha}|\Psi_0\rangle=0$, the state $|\Psi_0\rangle$ is often referred to as the Bogoliubov vacuum. 
It can be defined as $|\Psi_0\rangle\propto \prod_{p}(1+\lambda_{p\overline{p}}\hat{d}_{p}^\dagger \hat{d}_{\overline{p}}^\dagger)|\emptyset\rangle$ with particle pairs occupying $\hat{d}_{p}^\dagger \hat{d}_{\overline{p}}^\dagger|\emptyset\rangle$, where $\emptyset$ is a true vacuum.
Note that $\hat{d}_{r}^\dagger=\sum_{i}x_{ir}\hat{c}_{i}^\dagger$, where $r=p,\overline{p}$ are determined by a unitary transformation $x$ that brings $z=-(u^\dagger)^{-1}v^\dagger$ to a standard canonical form $\lambda$, i.e., $z=x\lambda x^\text{T}$. The only non-zero elements of $\lambda$ are $\lambda_{p\overline{p}}=-\lambda_{\overline{p}p}$ with $p$ and $\overline{p}$ being neighbouring odd and even numbers, respectively (more details can be found in Ref.~\cite{Mbeng_2020}; see also Ref.~\cite{Youla_1961}).

The above considerations can more formally be understood using the following symmetry property of $\hat H$~\cite{Kallin_2016,Liu_2017},
\begin{equation} \label{eq:symmetryprop}
    \Gamma\mathcal{H}\Gamma^{-1}=-\mathcal{H}\;,\;\;\;\;\;\;
    \Gamma=\sigma_x\mathcal{K}\;,
\end{equation}
where $\sigma_{x}$ is a generalization of the first Pauli matrix to a $2V \times 2V$ matrix with $V\times V$ identity matrices replacing 1, and $\mathcal{K}$ is the complex conjugation. The symmetry operator $\Gamma$ is antiunitary with $\Gamma^{-1}=\mathcal{K}\sigma_x$ and $\Gamma^2=1$. It corresponds to the particle-hole conjugation. 
Then, Eq.~(\ref{eq:symmetryprop}) implies
\begin{equation}
    \mathcal{H}\Gamma \psi_\alpha = -\Gamma\mathcal{H} \psi_\alpha =-\epsilon_\alpha\Gamma \psi_\alpha \;,
\end{equation}
and hence $\psi_{\alpha+V} = \Gamma \psi_\alpha$.
For simplicity, we will call both $\psi_\alpha$ and $\psi_{\alpha+V}$ as single-quasiparticle excitations (or single-quasiparticle eigenstates) in the rest of the paper.

It is often convenient to define the so-called charge~\cite{Ben_Shach_2015,Dominguez_2017}, which is for quasiparticle eigenstates $\psi_\alpha$ at $\alpha\leq V$ defined as
\begin{equation} \label{eq:def_charge}
    q_\alpha = \sum_{i=1}^{V}(|u_{i\alpha}|^2-|v_{i\alpha}|^2)\;,
\end{equation}
and has the same absolute value but opposite sign for $\Gamma\psi_{\alpha}$.
Furthermore, $q_\alpha=1$ for a pure particle excitation and $q_\alpha=-1$ for a pure hole excitation, while $q_\alpha=0$ for a perfect superposition of both. 
There is no simple relation between the sign of charge  $q_\alpha$ and the sign of energy $\epsilon_\alpha$.

The zero modes of $\mathcal{H}$, when emerge, require a special attention. If $\psi_\alpha$ is a zero mode, then $\Gamma\psi_{\alpha}$ is also a zero mode. One can built their linear combinations $\varphi_\beta=(\psi_\alpha+\Gamma\psi_\alpha)/\sqrt{2}$ and $\varphi_{\beta'}=(\psi_\alpha-\Gamma\psi_\alpha)/\sqrt{2}$~\cite{Gurarie_2007}, which both have zero charge, $q_\beta=q_{\beta'}=0$.
They are also eigenstates of the symmetry operator $\Gamma\varphi_\beta=\varphi_\beta$ and 
$\Gamma\varphi_{\beta'}=-\varphi_{\beta'}$, which resembles the property that the "particle" is its own "antiparticle".

\subsection{Quantum-chaotic quadratic model} \label{sec:ourmodel}

We now focus on the specific model that we study in the remainder of the paper.
While many studies of the Hamiltonians $\hat H$ of the form given by Eq.~(\ref{eqH}) aim at exploring their possible topological properties~\cite{Schynder_2008, Chiu_2016, Ludwig_2016}, we here pursue a different goal, namely, we set the parameters of $\hat H$ such that the model exhibits single-particle quantum chaos.
Specifically, we consider a cubic lattice with periodic boundary conditions and define $\hat H$ as a local Hamiltonian, also denoted as extended 3D Anderson model in Ref.~\cite{lydzba_rigol_21},
\begin{equation}
\label{eqHAnd}
\hat{H}=\sum_{\langle i,j \rangle} \left(- \hat{c}_i^\dagger \hat{c}_j +2 \Delta \hat{c}_i^\dagger \hat{c}_j^\dagger + {\rm h.c.} \right) + \sum_{i=1}^{V} \varepsilon_i  \hat{c}_i^\dagger \hat{c}_i \,,
\end{equation}
where $\langle i,j\rangle$ denote nearest neighbors.
The first term describes the hopping and pairing that is consistent with a spin-polarized triplet p-wave pairing interaction~\cite{Kitaev_2001, Kallin_2016, Liu_2017}.
The second term describes the on-site potential $\varepsilon_i = \frac{W}{2} r_i$ with a disorder strength $W$, where $r_i$ is a random number drawn independently from a uniform distribution over $[-1,1]$.

Since all model parameters in Eq.~(\ref{eqHAnd}) are real, it follows that $\mathcal{H}=\mathcal{H}^\text{T}$, the matrix $\mathcal{H}$ from Eq.~$\left(\ref{eqH2}\right)$ has a time-reversal symmetry, and the unitary matrix $U$ from Eq.~(\ref{eq:def_U}) is real.
The model belongs to the BDI symmetry class~\cite{Schynder_2008,Zirnbauer_2010,Chiu_2016}. In the limit $\Delta\rightarrow 0$, the 3D Anderson model from the AI symmetry class~\cite{Schynder_2008,Zirnbauer_2010,Chiu_2016} is recovered~\cite{Anderson_58}.

\begin{figure}[t!]
\centering
\includegraphics[width=0.9\columnwidth]{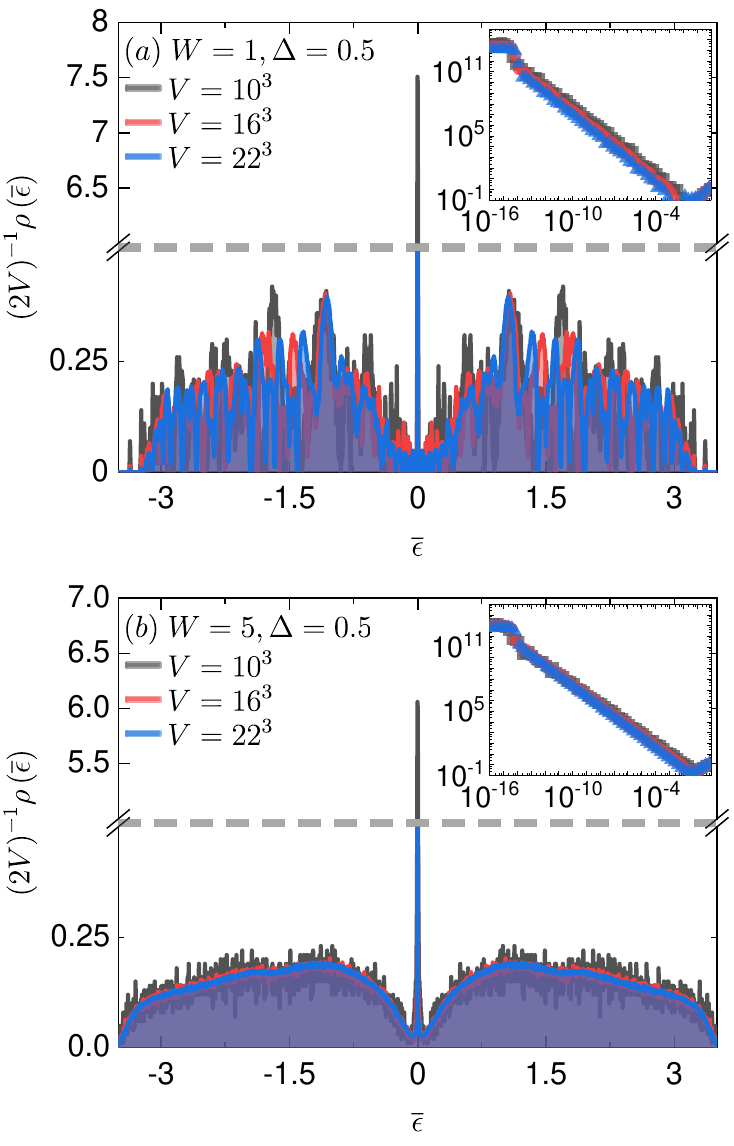}
\caption{The scaled density of single-quasiparticle eigenstates $(2V)^{-1} \rho\left(\overline{\epsilon}\right)$, see text for a definition, at $\Delta=0.5$.
Results are shown for the disorder strengths (a) $W=1$ and (b) $W=5$, for three system sizes $V=10^3, 16^3, 22^3$, and averaged over $10$ disorder realizations. The bin size is $\delta\overline{\epsilon}=5\cdot 10^{-3}.$
The insets display the same results as the main panel but with logarithmic scales on vertical and horizontal axes. Here, the binning is performed in the logarithm of energy, so that the~bin size is $\delta\log\left(\overline{\epsilon}\right)=0.25$.
}
\label{fig3}
\end{figure}

We observe that the model in Eq.~(\ref{eqHAnd}) exhibits disorder-induced near zero modes at nonzero $\Delta$.
Figure~\ref{fig3} shows the scaled density $(2V)^{-1} \rho\left(\overline{\epsilon}\right)$, where the density of single-quasiparticle eigenstates is
$\rho\left(\overline{\epsilon}\right) = \delta N / \delta \epsilon |_{\overline{\epsilon}}$.
Its most prominent feature is a sharp peak at $\epsilon = 0$, which denotes the near zero modes.
The peak is present at $\Delta=0.5$ at both weak disorder [$W=1$, Fig.~\ref{fig3}(a)] and moderate disorder [$W=5$, Fig.~\ref{fig3}(b)], while it is absent at $\Delta=0$ as well as at weak $\Delta$ such as $\Delta=0.1$ (not shown).
Results in the insets of Figs.~\ref{fig3}(a) and~\ref{fig3}(b) show a polynomial divergence at zero energy that is independent of the system size~$V$ and is well described by a function $\epsilon^{-\zeta}$, with $\zeta\in[0.95,1.1]$.What is not visible in the logarithmic scale is that the height of the peak slowly decreases with the system size~$V$.
A more quantitative analysis of the near zero modes will be carried out in Sec.~\ref{sec:zeromodes}.

\section{Quasiparticle eigenstates} \label{sec:wavefunction}

Eigenstate thermalization is expected to occur in eigenstates that are sufficiently delocalized in a non fine-tuned basis.
The goal of this section is to detect the regime of parameters $W$ and $\Delta$, for which the statistical properties of the majority of single-quasiparticle eigenstates are consistent with the predictions of quantum chaos.

\subsection{Inverse participation ratio} \label{sec:ipr}

We calculate the inverse participation ratio (IPR) of a single-quasiparticle eigenstate, which we define as
\begin{equation} \label{eq:def_ipralpha}
\text{IPR}_\alpha = \sum_{i=1}^{V} \left( u_{i\alpha}^4 + v_{i\alpha}^4 \right)\;, 
\end{equation}
and average it throughout the single-quasiparticle spectrum,
\begin{equation} \label{eq:def_iprav}
\text{IPR}_\text{av}= \frac{1}{2V}\sum_{\alpha=1}^{2V} \text{IPR}_\alpha \;.
\end{equation}
The above definitions have been previously used for systems with superconducting pairing terms~\cite{Wang_2016,Yahyavi_2019} (see also~\cite{Fraxanet_2022} for a different definition).

\begin{figure}[t!]
\centering
\includegraphics[width=\columnwidth]{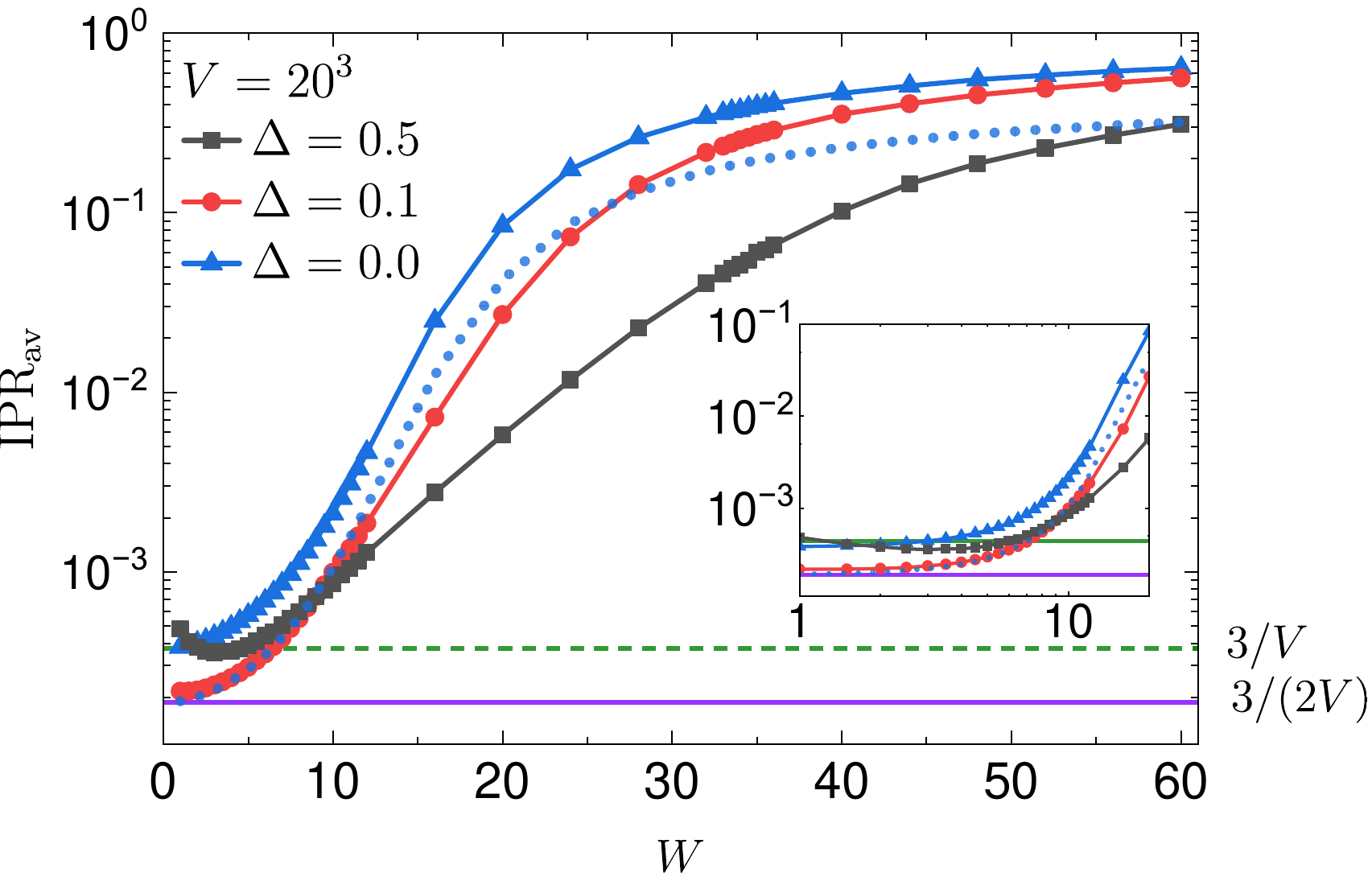}
\caption{$\text{IPR}_\text{av}$ from Eq.~(\ref{eq:def_iprav}) versus disorder strength $W$ at $V=20^3$ and $\Delta=0, 0.1, 0.5$.
Results are averaged over $10$ disorder realizations.
Horizontal dashed and solid lines denote the GOE predictions $\text{IPR}_\text{GOE}=3/V$ and $3/(2V)$ that apply to the cases $\Delta=0$ and $\Delta\neq 0$, respectively.
The blue dotted line corresponds to $\text{IPR}_\text{av}/2$ at $\Delta=0$.
The inset displays the same results using a logarithmic scale on the horizontal axis.
}
\label{fig1}
\end{figure}

In single-quasiparticle eigenstates of quantum-chaotic quadratic Hamiltonians at $\Delta\neq 0$, the amplitudes $u_{i\alpha}$ and $v_{i\alpha}$ are expected to be (pseudo)random numbers drawn from a normal distribution with zero mean and $\frac{1}{2V}$ variance.
In this case, the IPR is expected to agree with the Gaussian orthogonal ensemble (GOE) prediction $\text{IPR}_\text{GOE}=\frac{3}{2V}$~\cite{ZELEVINSKY_1996}.
At the particle number conserving point $\Delta=0$, the GOE prediction is $\text{IPR}_\text{GOE}=\frac{3}{V}$.
If, in contrast, the eigenstates are localized in some basis, the IPR in this basis is a nonzero constant that is independent of the system size $V$.
In the model from Eq.~(\ref{eqHAnd}), the localization is expected to occur in the site-occupation basis at large disorder~$W$.
In this paper, we are interested in the quantum-chaotic regime of the model from Eq.~(\ref{eqHAnd}), and hence we only focus on the weak disorder regime.

Figure~\ref{fig1} shows $\text{IPR}_\text{av}$ versus disorder strength~$W$ at three different values of $\Delta=0, 0.1, 0.5$.
In all cases $\text{IPR}_\text{av}$ is very small at weak disorder $W$, indicating delocalization is site-occupation basis, and it increases at large $W$, indicating localization.
A more detailed view into the weak disorder regime, see also the inset of Fig.~\ref{fig1}, reveals that $\text{IPR}_\text{av}$ approaches the corresponding GOE prediction at $\Delta=0$ and 0.1.
At $\Delta=0.5$, however, $\text{IPR}_\text{av}$ shows a non-monotonous behavior and, at least at $V=20^3$, does not entirely match with the GOE prediction.

\begin{figure}[t!]
\centering
\includegraphics[width=\columnwidth]{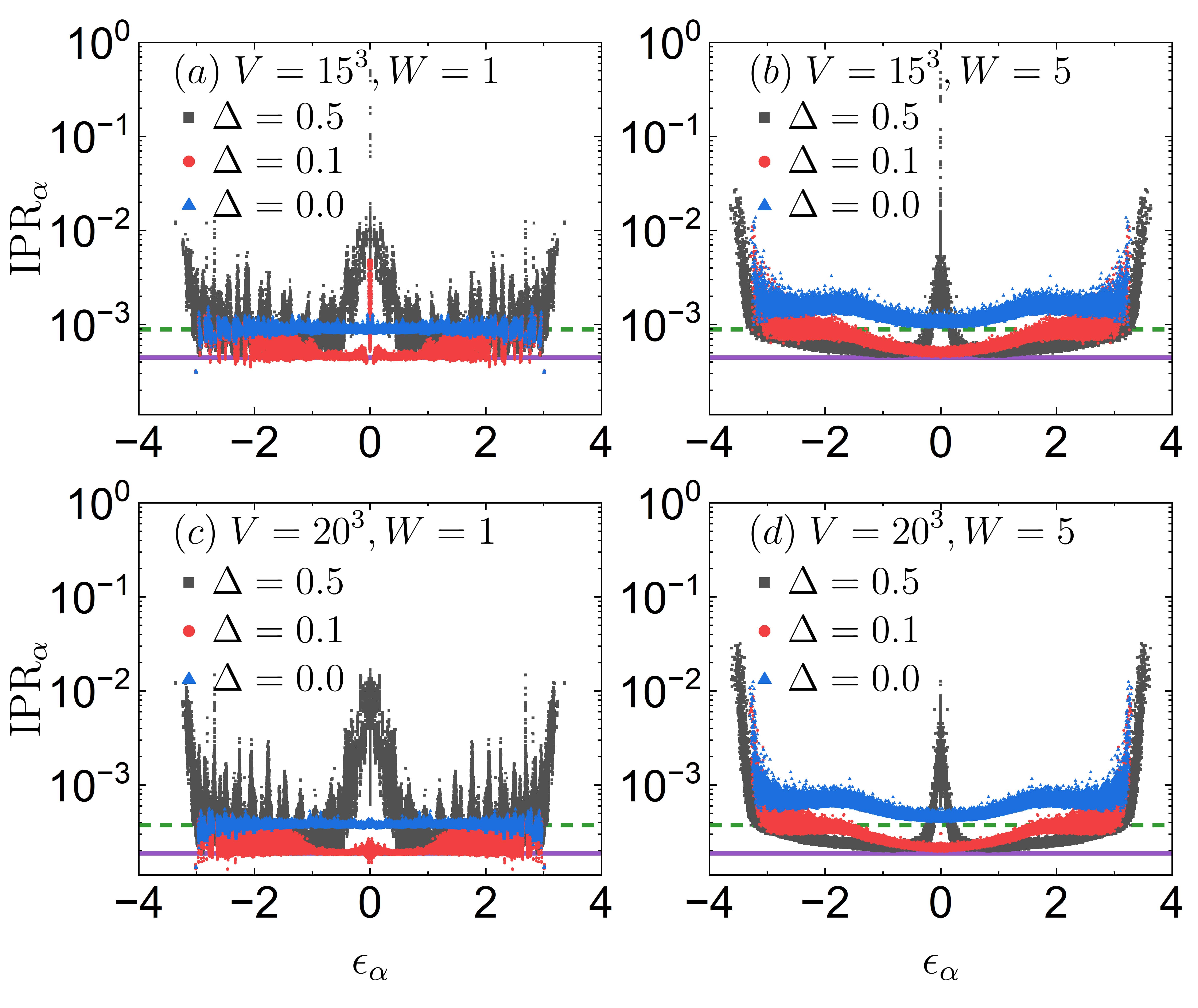}
\caption{$\text{IPR}_\alpha$ from Eq.~(\ref{eq:def_ipralpha}) versus quasiparticle energy $\epsilon_\alpha$ at $\Delta=0, 0.1, 0.5$.
Results are presented for $10$ disorder realizations.
(a) $W=1$ and $V=15^3$, (b) $W=5$ and $V=15^3$, (c) $W=1$ and $V=20^3$, (d) $W=5$ and $V=20^3$.
Horizontal dashed and solid lines denote the GOE predictions $\text{IPR}_\text{GOE}=3/V$ and $3/(2V)$ that apply to the cases $\Delta=0$ and $\Delta\neq 0$, respectively.}
\label{fig2}
\end{figure}

To better understand the small deviations from the GOE prediction at $\Delta=0.5$, we study ${\rm IPR}_\alpha$ versus quasiparticle energy $\epsilon_\alpha$ in Fig.~\ref{fig2}, focusing on disorders $W=1$ and 5.
Results at $\Delta=0$ and 0.1, which are also included in Fig.~\ref{fig2}, are rather expected: the IPR's match quite accurately with the GOE predictions at $W=1$, while the agreement deteriorates at $W=5$.
In the latter case, the deviations from the GOE predictions are most prominent at spectral edges, and may indicate a tendency towards forming a mobility edge.

On the contrary, the numerical results at $\Delta=0.5$ in Fig.~\ref{fig2} show a large deviation of $\text{IPR}_\alpha$ from the GOE prediction at $|\epsilon_\alpha| \approx 0$.
In this energy regime, near zero modes were observed in the density of states in Fig.~\ref{fig3}.
The anomalous behavior of $\text{IPR}_\alpha$ at $|\epsilon_\alpha| \approx 0$ is most clear at $W=5$, see Figs.~\ref{fig2}(b) and~\ref{fig2}(d), while at $W=1$ other subbands emerge at $|\epsilon_\alpha| \neq 0$, see Figs.~\ref{fig2}(a) and~\ref{fig2}(c).
Moreover, even at $\Delta=0.1$ and $V=15^3$, see Fig.~\ref{fig2}(a), one observes deviations of $\text{IPR}_\alpha$ from the GOE prediction at $|\epsilon_\alpha| \approx 0$, which however disappears at larger system size $V=20^2$, see Fig.~\ref{fig2}(c).
These observations call for further quantitative studies of properties of near zero modes, which we carry out in the next section.

\subsection{Disorder-induced near zero modes} \label{sec:zeromodes}

We now study the structure of disorder-induced near zero modes.
As a working definition, the term {\it near zero modes} refers to quasiparticle eigenstates with small absolute energy $\epsilon_\alpha$ that is still sufficiently larger than the machine precision, $10^{-14} \lesssim |\epsilon_\alpha| \ll 1$, while with {\it zero modes} we have in mind the quasiparticle eigenstates with zero energy within the machine precision, $|\epsilon_\alpha| \lesssim 10^{-14}$.
We observe that the near zero modes can be organized in pairs $(\alpha,\alpha')$ with $\epsilon_{\alpha'} \approx -\epsilon_{\alpha}$ and $q_{\alpha'} \approx -q_{\alpha}$, with negligible relative differences.
In these pairs, $u_{i\alpha'}=v_{i\alpha}$ and $v_{i\alpha'}=u_{i\alpha}$ for the majority of sites $i$.
This organisation is impossible for most of the zero modes.

We calculate the expectation values of two operators, the particle-hole conjugation  $|\langle{\Gamma}\rangle_\alpha|=|\langle\alpha|{\Gamma}|\alpha\rangle|$ and the charge $|q_\alpha|$. 
In this and all further sections, we use the bra-ket notation in which $|\alpha\rangle$ denotes the single-quasiparticle eigenstate.
One expects $|\langle{\Gamma}\rangle_\alpha| \approx 0$ and $|q_\alpha|>0$ for states sufficiently away from zero modes (i.e., away from zero quasiparticle energy), while for zero modes one expects $|\langle{\Gamma}\rangle_\alpha| = 1$ and $|q_\alpha|=0$.
A way to consider these properties is to say that the quasiparticle eigenstates with a nonvanishing $|\langle{\Gamma}\rangle_\alpha|$ are not particle-hole symmetric, since $\Gamma|\alpha\rangle$ is not orthogonal to $|\alpha\rangle$.
Therefore, $|\langle{\Gamma}\rangle_\alpha|$ measures the deviation from the particle-hole symmetry.

\begin{figure}[t!]
\centering
\includegraphics[width=\columnwidth]{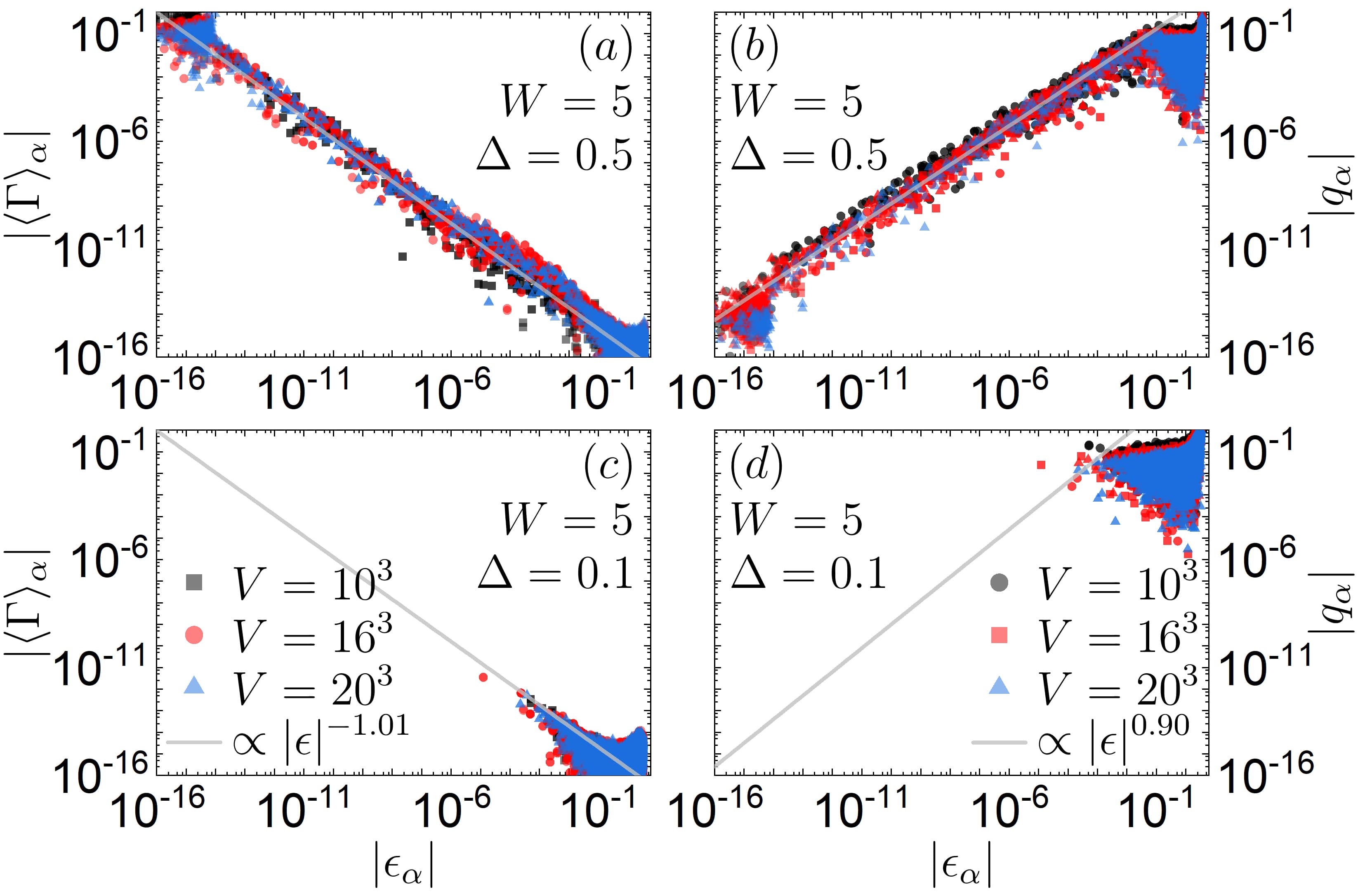}
\caption{(a,c) Particle-hole conjugation $|\langle{\Gamma}\rangle_\alpha|$ and (b,d) charge $|q_\alpha|$ versus energy $|\epsilon_\alpha|$, at $W=5$.
(a,b) $\Delta=0.5$ and (c,d) $\Delta = 0.1$.
We present $5, 3$ and $1$ disorder realizations for $V=10^3, 16^3$ and $20^3$, respectively.
Lines are fits of the functions $a_1|\epsilon_\alpha|^{-\zeta}$ to the results in (a) and $a_2|\epsilon_\alpha|^{\zeta}$ to the results in (b).}
\label{fig4}
\end{figure}

\begin{figure}[t!]
\centering
\includegraphics[width=\columnwidth]{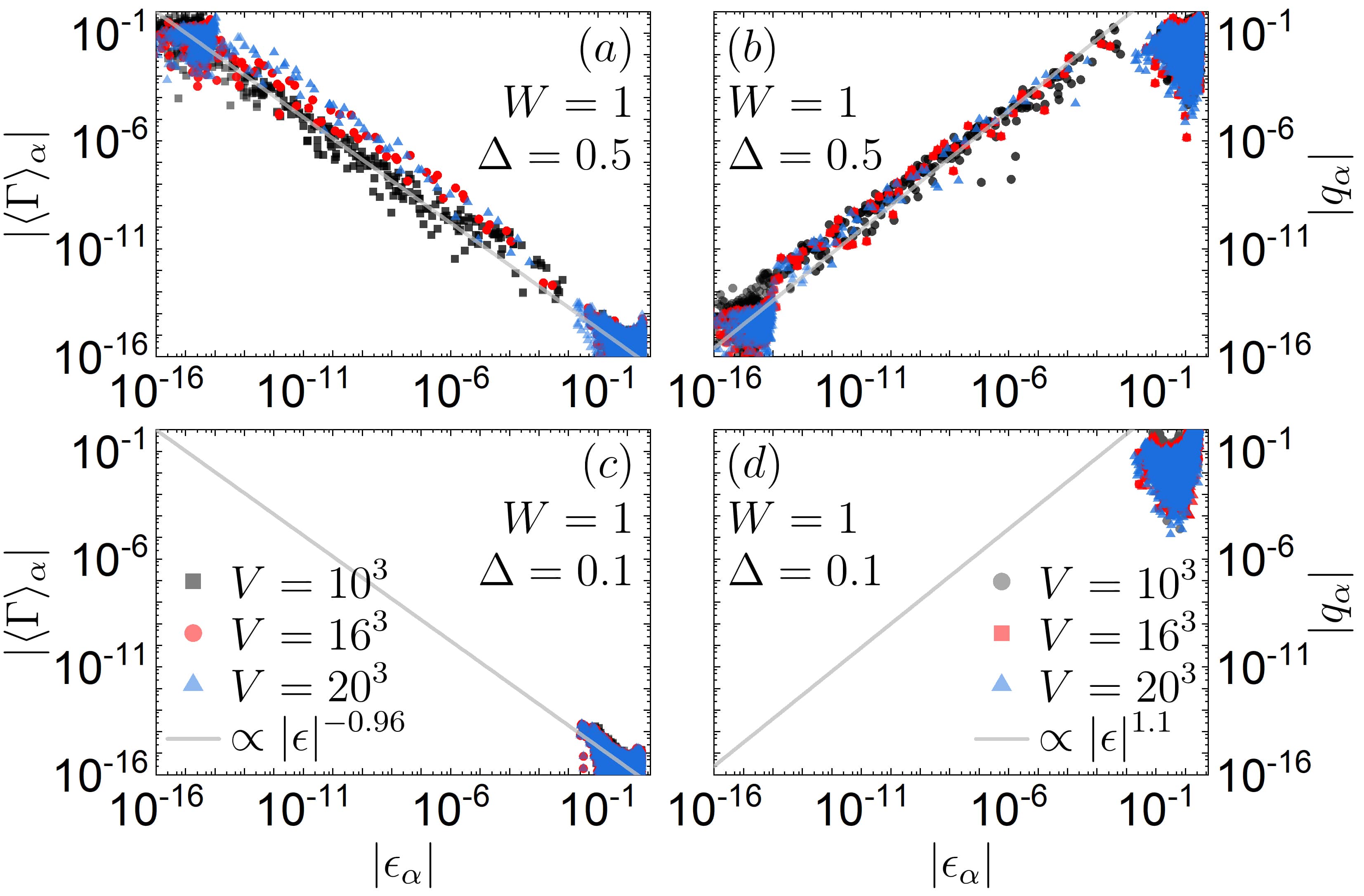}
\caption{(a,c) Particle-hole conjugation $|\langle{\Gamma}\rangle_\alpha|$ and (b,d) charge $|q_\alpha|$ versus energy $|\epsilon_\alpha|$, at $W=1$.
(a,b) $\Delta=0.5$ and (c,d) $\Delta = 0.1$.
We present $5, 3$ and $1$ disorder realizations for $V=10^3, 16^3$ and $20^3$, respectively.
Lines are fits of the functions $a_1|\epsilon_\alpha|^{-\zeta}$ to the results in (a) and $a_2|\epsilon_\alpha|^{\zeta}$ to the results in (b).}
\label{fig5}
\end{figure}

Figure~\ref{fig4} shows results for $|\langle{\Gamma}\rangle_\alpha|$ and $|q_\alpha|$ at $W=5$, while the analogous results at $W=1$ are shown in Fig.~\ref{fig5}.
There are two main results.
The first is that at $\Delta=0.1$, for which the IPR results in Sec.~\ref{sec:ipr} matched with the GOE prediction to a reasonable accuracy, we observe $|\langle{\Gamma}\rangle_\alpha| \approx 0$ and $|q_\alpha|>0$ for all single-quasiparticle eigenstates.
This results is consistent with absence of (near) zero modes at $\Delta=0.1$.
The second result is that at $\Delta=0.5$, the values of $|\langle{\Gamma}\rangle_\alpha|$ and $|q_\alpha|$ smoothly span over the entire interval from 0 to 1.
They appear to be a well-defined function of the absolute energy $|\epsilon_\alpha|$ and roughly independent of the system size $V$.
We obtain a good fit using $|\langle{\Gamma}\rangle_\alpha| \propto |\epsilon_\alpha|^{-\zeta}$ and  $|q_\alpha| \propto |\epsilon_\alpha|^{\zeta}$, where $\zeta \approx 1$ in both cases.
This suggest emergence of a smooth crossover between zero modes and states than cannot be characterized as zero modes, and it reinforces our term {\it near zero modes} to describe the states at small but still nonzero energy.

\begin{figure}[t!]
\centering
\includegraphics[width=0.85\columnwidth]{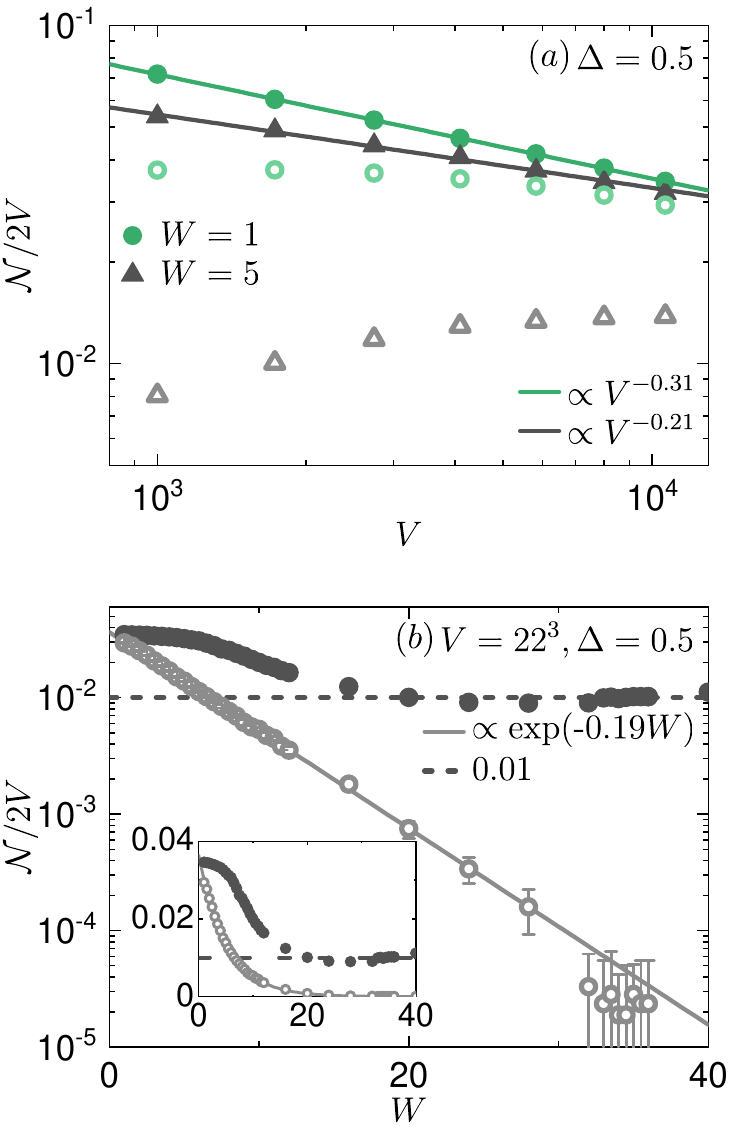}
\caption{
Scaling of the fraction of zero and near zero modes $\mathcal{N}/(2V)$ at $\Delta=0.5$.
(a) Scaling with system size $V$ at $W=1$ and 5.
(b) Scaling with disorder $W$ at $V=22^3$.
Results are averaged over $50$ disorder realizations, and error bars correspond to standard deviations with respect to disorder realizations.
Open symbols are results using a zero mode condition, while filled symbols are results using a near zero mode condition (see text for details).
Lines are fits to the data of the functions: (a) $b_1 V^\zeta$, where $\zeta=0.31$ at $W=1$ and $\zeta=0.21$ at $W=5$, and (b) $b_2\exp(-\zeta W)$, where $\zeta=0.19$. }
\label{fig6}
\end{figure}

We next determine the number of zero and near zero modes $\mathcal{N}$, and study their scaling with the total number of states $2V$.
Note, however, that results in Figs.~\ref{fig4} and~\ref{fig5} hint at a certain degree of ambiguity when defining $\mathcal{N}$.
We consider two criteria.
(i) Zero mode condition: zero modes correspond to quasiparticle eigenstates with $\langle\hat{\Gamma}\rangle_\alpha>10^{-4}$ and $q_\alpha<10^{-14}$.
According to the results in Figs.~\ref{fig4} and~\ref{fig5}, zero modes can be interpreted as those with energy that is zero within machine precision, which is consistent with our working definition of zero modes introduced above.
(ii) Near zero mode condition: near zero modes correspond to quasiparticle eigenstates with $\langle\hat{\Gamma}\rangle_\alpha>10^{-14}$ and $q_\alpha<10^{-4}$.
Focusing on $\Delta=0.5$, the scaling properties of the relative number of zero and near zero modes $\mathcal{N}/(2V)$ are shown in Fig.~\ref{fig6}.

Figure~\ref{fig6}(a) shows the scaling of $\mathcal{N}/(2V)$ versus $V$ at disorders $W=1$ and 5.
Results using the near zero mode condition exhibit a rather clear polynomial decay $V^{-\zeta}$, with $\zeta=0.31$ at $W=1$ and $\zeta=0.21$ at $W=5$.
This suggests that the number of near zero modes is subextensive, and it eventually represents a vanishing fraction of all single-quasiparticle eigenstates in the thermodynamic limit.
Results using the zero mode condition, on the other hand, do not exhibit a clear decay, in particular at $W=5$.
However, they are upper bounded by the decay given by the near zero mode condition. Therefore, one may expect that the fraction of zero modes decays to zero irrespective of the precise quantitative criterion for their definition.

Figure~\ref{fig6}(b), in contrast, shows the scaling of $\mathcal{N}/(2V)$ versus disorder $W$ at a fixed system size $V=22^3$.
In this case, results using the zero mode condition exhibit an exponential decay $\exp(-\zeta W)$, with $\zeta=0.19$ at $V=22^3$.
This suggests that the contribution of zero modes is negligible at moderate and large $W$.
On the other hand, results using the near zero mode condition saturate to a constant at fixed $V$.
This results is expected from Fig.~\ref{fig6}(a), which suggest that the fraction $\mathcal{N}/(2V)$ is nonzero for any finite $V$.

\section{Single-quasiparticle eigenstate thermalization} \label{sec:sqETH}

We now turn our attention to the main topic of the paper, i.e., eigenstate thermalization in single-quasiparticle eigenstates of the Hamiltonian from Eq.~(\ref{eqHAnd}).
Based on the analysis in Sec.~\ref{sec:wavefunction}, we focus on the disorder strength $W=5$ and two distinct pairing strengths $\Delta=0.1$ and $0.5$. 
As shown in Figs.~\ref{fig4} and~\ref{fig5}, the system exhibits zero and near zero modes at $\Delta=0.5$ but not at $\Delta=0.1$, at least for the system sizes under investigation.

We conjecture that the matrix elements of a normalized observable $\underline{\hat{O}}$ (to be defined in Sec.~\ref{sec:normalization}) in single-quasiparticle eigenstates of quantum-chaotic quadratic Hamiltonians, away from zero modes, can be written as
\begin{equation}
\label{eqETH}
    \langle\alpha|\underline{\hat{O}}|\beta\rangle = \mathcal{O}(\overline{\epsilon})\delta_{\alpha\beta}+
    \rho(\overline{\epsilon})^{-1/2}\mathcal{F}(\overline{\epsilon},\omega)R_{\alpha\beta}\;,
\end{equation}
where $\mathcal{O}(\overline{\epsilon})$ and
$\mathcal{F}(\overline{\epsilon},\omega)$ are smooth functions of their arguments, and $R_{\alpha\beta}$ is a random variable with zero mean and unit variance. 
In Eq.~(\ref{eqETH}), $|\alpha\rangle$ and $|\beta\rangle$ represent single-quasiparticle eigenstates, and hence $\overline{\epsilon} = (\epsilon_\alpha + \epsilon_\beta)/2$ and $\omega =\epsilon_\alpha-\epsilon_\beta$ refer to the mean single-quasiparticle energy and the corresponding difference, respectively. 
The fluctuations of matrix elements are governed by the single-quasiparticle density of states
$\rho(\overline{\epsilon}) = \delta N / \delta \epsilon |_{\overline{\epsilon}}$
studied in Fig.~\ref{fig3}, which scales with the system size $V$. 
The ansatz in Eq.~(\ref{eqETH}) carries some analogies, but also differences, with respect to the eigenstate-thermalization hypothesis (ETH) ansatz for many-body eigenstates of quantum-chaotic interacting Hamiltonians~\cite{deutsch_91, srednicki_94, rigol_dunjko_08, dalessio_kafri_16}.
To highlight the distinction between interacting and quadratic systems, we refer to the ansatz in Eq.~(\ref{eqETH}) as the single-quasiparticle eigenstate thermalization ansatz.

In a system with $V$ lattice sites, we consider $V$ distinct single-quasiparticle eigenstates $|\alpha\rangle$.
Since the Hamiltonian $\hat H$ from Eq.~(\ref{eqH2}) contains $2V$ eigenstates, there may exist a certain degree of ambiguity about the choice of a single-quasiparticle eigenstate from a pair of eigenstates with energies $\pm \epsilon_\alpha$ related by a particle-hole symmetry~$\Gamma$.
One possibility is to consider single-quasiparticle eigenstates with non-negative energies $\epsilon_\alpha\geq 0$. 
In this case, one can define $|\alpha\rangle = \hat f_\alpha^\dagger |\Psi_0\rangle$, where the "vacuum" state $|\Psi_0\rangle$ is the many-body ground state (see also the discussion in Sec.~\ref{sec:model}).
However, to make a connection with the single-particle eigenstate thermalization in the 3D Anderson model from Ref.~\cite{lydzba_rigol_21} (the limit $\Delta\rightarrow0$), we define the "vacuum" state $|\Theta_0\rangle$ as the state that is annihilated by the quasiparticles with negative charge $q_\alpha$, which energy is $E_{\Theta_0}=\sum_{q_\alpha<0}\epsilon_\alpha\approx 0$.
Hence, $|\alpha\rangle$ refers to the single-quasiparticle eigenstate of the form $\hat f_\alpha^\dagger|\Theta_0\rangle$ with energy $E_{\Theta_0}+2\epsilon_\alpha\approx 2\epsilon_\alpha$, where $\hat f_\alpha^\dagger$ creates a quasiparticle with a positive charge~$q_\alpha$.
Nevertheless, we do not expect this choice to have an impact on the validity of the single-quasiparticle eigenstate thermalization.

\subsection{Observable normalization} \label{sec:normalization}

The observable $\underline{\hat{O}}$ considered in Eq.~(\ref{eqETH}) is labeled by the underlined letter, which means that it is normalized and traceless.
The trace is carried out in the single-quasiparticle Hilbert space, and hence tracelessness requires
\begin{equation}
    \text{Tr} \left\{\underline{\hat{O}}\right\} = \sum_{\alpha=1}^V \langle\alpha| \underline{\hat{O}} | \alpha\rangle = 0 \;,
\end{equation}
while normalization is defined by a unit Hilbert-Schmidt norm,
\begin{equation} \label{eq:norm}
    ||\underline{\hat{O}}||^2=\frac{1}{V}\text{Tr} \left\{\underline{\hat{O}}^2\right\} = \frac{1}{V} \sum_{\alpha,\beta=1}^V |\langle\alpha| \underline{\hat{O}} | \beta\rangle|^2 =1.
\end{equation}
As a side remark, we note that normalization of an observable $\underline{\hat{O}}$ in the single-quasiparticle Hilbert space does not imply normalization of this observable in a many-body Hilbert space of dimension $2^V$, for which $1/V$ in Eq.~(\ref{eq:norm}) is replaced by $1/2^V$ and the trace runs over $2^V$ many-body states.

The observable for which we carry out the numerical calculations to test Eq.~(\ref{eqETH}) is the site occupation of the lattice site $i$,
\begin{equation}
\hat{n}_i=\hat{c}_{i}^\dagger \hat{c}_{i}\;.
\end{equation}
The matrix elements of the observable are obtained by expressing the bare spinless fermion creation and annihilation operators $\hat c_i^\dagger, \hat c_i$ by the quasiparticle operators $\hat f_\alpha^\dagger, \hat f_\alpha$ from Eq.~(\ref{eq:def_falpha}).
The diagonal matrix elements are
\begin{equation} \label{eq:def_naa}
    \langle\alpha|\hat{n}_{i}|\alpha\rangle=u_{i\alpha}^2-v_{i\alpha}^2+\sum_{\beta=1}^{V} v_{i\beta}^2\;,
\end{equation}
while offdiagonal matrix elements are
\begin{equation} \label{eq:def_nab}
    \langle\alpha|\hat{n}_{i}|\beta\rangle=u_{i\alpha}u_{i\beta}-v_{i\alpha}v_{i\beta}\;.
\end{equation}
The observable $\hat{n}_i$ has an $O(1)$ Hilbert-Schmidt norm in the many-body Hilbert space.
However, the traceless and normalized counterpart of $\hat n_i$ in the single-quasiparticle Hilbert space, our space of interest, is
\begin{equation} \label{eq:def_n_normalized}
\underline{\hat{n}}_i=\frac{1}{\sqrt{\text{N}}}\left(\hat{n}_i-\text{T}\right) \;,
\end{equation}
where
\begin{equation} \label{eq:def_T}
\text{T}=\frac{1}{V}\sum_{\alpha=1}^{V}(u_{i\alpha}^2-v_{i\alpha}^2)+\sum_{\alpha=1}^{V}v_{i\alpha}^2
\end{equation}
and
\begin{align}
\label{eqnorm}
\text{N} & = \frac{1}{V}\left(\sum_{\alpha=1}^{V}u_{i\alpha}^2\right)^2
+\frac{V-1}{V}\left(\sum_{\alpha=1}^{V}v_{i\alpha}^2\right)^2 \nonumber \\
&+\frac{2}{V}\sum_{\alpha=1}^{V}u_{i\alpha}^2\sum_{\alpha=1}^{V}v_{i\alpha}^2
-\frac{2}{V}\left(\sum_{\alpha=1}^{V}u_{i\alpha}v_{i\alpha}\right)^2
-\text{T}^2\;.
\end{align}
The magnitude of $\rm T$ from Eq.~(\ref{eq:def_T}) is ${\rm T} \approx 1/2$, which is a consequence of the many-body content of $|\alpha\rangle$ in terms of bare fermions.
This property differs from the particle number conserving quadratic systems in which ${\rm T} \propto 1/V$~\cite{lydzba_zhang_21}.
In contrast, the magnitude of $\rm N$ from Eq.~(\ref{eqnorm}) is ${\rm N} \propto 1/V$, which is a consequence of the restriction to the single-quasiparticle subspace.
This property is analogous to the particle number conserving quadratic systems.

Without the loss of generality, we fix $i=1$ and replace $\hat{n}_{1}\rightarrow \hat{n}$ ($\underline{\hat{n}}_{1}\rightarrow \underline{\hat{n}}$) in what follows. To simplify the notation, we define $n_{\alpha\alpha}=\langle\alpha|\hat{n}|\alpha\rangle$ ($\underline{n}_{\alpha\alpha}=\langle\alpha|\underline{\hat{n}}|\alpha\rangle$) and $n_{\alpha\beta}=\langle\alpha|\hat{n}|\beta\rangle$ ($\underline{n}_{\alpha\beta}=\langle\alpha|\underline{\hat{n}}|\beta\rangle$). 

\subsection{Structure of matrix elements} \label{sec:structure}

Next, we numerically test the validity of the ansatz in Eq.~(\ref{eqETH}) for the observable in Eq.~(\ref{eq:def_n_normalized}).
We first focus on the structure functions $\mathcal{O}(\overline{\epsilon})$ and $\mathcal{F}(\overline{\epsilon},\omega)$, and comment on the amplitudes of the matrix elements of near zero modes at $\Delta=0.5$.

\subsubsection{Diagonal matrix elements}

\begin{figure}[t!]
\centering
\includegraphics[width=0.85\columnwidth]{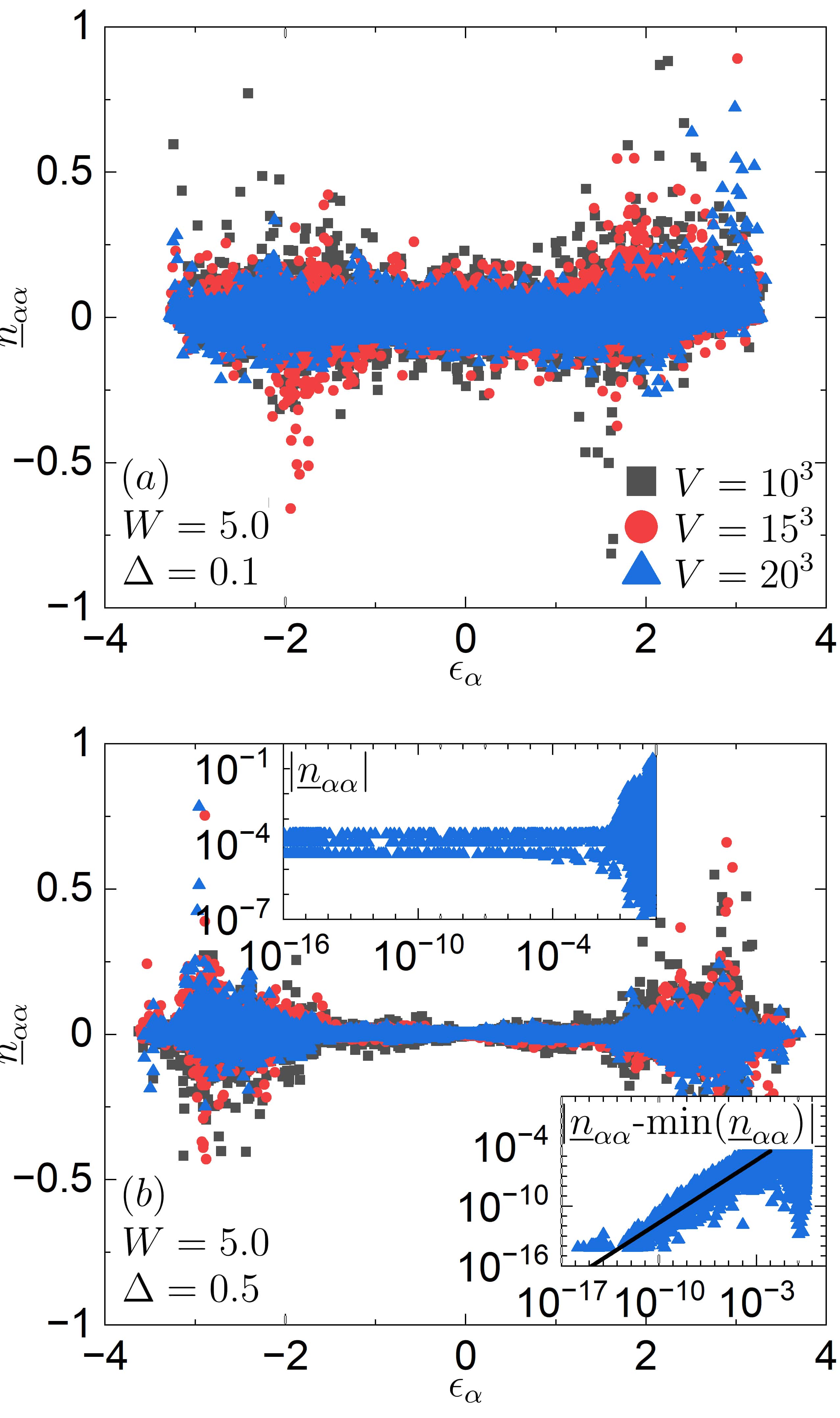}
\caption{Diagonal matrix elements $\underline{n}_{\alpha\alpha}$ as functions of eigenenergies $\epsilon_\alpha$ at disorder $W=5$ and the pairing strengths (a) $\Delta=0.1$ and (b) $\Delta=0.5$.
Results are shown for three system sizes $V\in\left\{10^3,15^3,20^3\right\}$ and $10$ disorder realizations.
Insets of (b): results at $\Delta=0.5$ and $V=20^3$, shown on a logarithmic scale on both axes.
The upper inset shows $|\underline{n}_{\alpha\alpha}|$ and the lower inset shows $|\underline{n}_{\alpha\alpha}-\text{min}(|\underline{n}_{\alpha\alpha}|)|$, where $\text{min}(|\underline{n}_{\alpha\alpha}|)$ is the minimal value of $|\underline{n}_{\alpha\alpha}|$ within the band of near zero modes with energies $\epsilon_\alpha\le 10^{-4}$.
The solid line in the lower inset denotes the function $\propto|\epsilon|^\zeta$, where $\zeta=0.90$ was obtained using the linear regression.
}
\label{figd1}
\end{figure}

The diagonal matrix elements $\underline{n}_{\alpha\alpha}$ versus $\epsilon_\alpha$ are shown for three system sizes $V\in\left\{10^3,15^3,20^3\right\}$ at $\Delta=0.1$ and $0.5$ in Figs.~\ref{figd1}(a) and~\ref{figd1}(b), respectively.
Results suggest that the matrix elements are structureless since $\underline{n}_{\alpha\alpha}$ fluctuate around zero.
Specifically, the unnormalized diagonal matrix elements of $\underline{n}_{\alpha\alpha}$ can be expressed using Eqs.~(\ref{eq:def_naa}) and~(\ref{eq:def_T}) as
\begin{equation} \label{eq:naa_T}
    n_{\alpha\alpha} - {\rm T} = u_{i\alpha}^2 - v_{i\alpha}^2 - \left( u_{i,\overline{\alpha}}^2 - v_{i,\overline{\alpha}}^2\right) \;,
\end{equation}
where the eigenstate-averaged coefficients $u_{i,\overline{\alpha}}^2$ and $v_{i,\overline{\alpha}}^2$ are defined as $u_{i,\overline{\alpha}}^2 = (1/V)\sum_{\alpha'=1}^V u_{i\alpha'}^2$ and $v_{i,\overline{\alpha}}^2 = (1/V)\sum_{\alpha'=1}^V v_{i\alpha'}^2$, respectively.
Using the notion of a {\it site-resolved} charge $u_{i\alpha}^2 - v_{i\alpha}^2$ [cf.~Eq.~(\ref{eq:def_charge})], the diagonal matrix elements of the traceless observable $\underline{n}_{\alpha\alpha}$ can be seen as a measure of fluctuations of a site-resolved charge over the averaged site-resolved charge $u_{i,\overline{\alpha}}^2 - v_{i,\overline{\alpha}}^2$.
The fluctuations of matrix elements appear to decrease with $V$ and will be studied in more detail in Sec.~\ref{sec_fluc}.
The support of fluctuations at $\Delta=0.1$ appears to be rather insensitive to $\epsilon_\alpha$, while at $\Delta=0.5$ it is larger at the edges of the single-quasiparticle spectrum.

Even though the results in Fig.~\ref{figd1} may suggest that $\mathcal{O}(\overline{\epsilon}) \approx 0$ in Eq.~(\ref{eqETH}), there nevertheless exists a fine structure of near zero modes that we study in the insets of Fig.~\ref{figd1}(b).
Figure~\ref{fig4} shows that the charge $q_\alpha$ of near zero modes vanishes as $\epsilon_\alpha\to 0$ and, hence, one may also expect the site-resolved charge $u_{i\alpha}^2 - v_{i\alpha}^2$ to vanish as well.
Equation~(\ref{eq:naa_T}) then suggest that the diagonal matrix elements of near zero modes become independent of the eigenstate index $\alpha$ since 
$n_{\alpha\alpha} - {\rm T} \approx - \left( u_{i,\overline{\alpha}}^2 - v_{i,\overline{\alpha}}^2\right)$.
The upper inset of Fig.~\ref{figd1}(b) shows that this is indeed the case.
However, in the lower inset of Fig.~\ref{figd1}(b) we show the subtracted matrix elements $|\underline{n}_{\alpha\alpha}-\text{min}(|\underline{n}_{\alpha\alpha}|)|$, where $\text{min}(|\underline{n}_{\alpha\alpha}|)$ is the minimal absolute value of matrix elements within the band of near zero modes.
These matrix elements exhibit the structure $|\underline{n}_{\alpha\alpha}-\text{min}(|\underline{n}_{\alpha\alpha}|)|\propto|\epsilon|^\zeta$ with $\zeta=0.90$, see the solid line in the lower inset of Fig.~\ref{figd1}(b).
This property is consistent with the dependence of charge $q_\alpha$ on $\epsilon_\alpha$, as shown in Fig.~\ref{fig4}.

\subsubsection{Offdiagonal matrix elements}

The offdiagonal matrix elements $\underline{n}_{\alpha\beta}$ are calculated for pairs of $\ket{\alpha}$ and $\ket{\beta}$ restricted to a narrow energy window $\delta$ around a target energy $\epsilon_\text{tar}$, i.e., $|\overline{\epsilon}-\epsilon_\text{tar}|<\delta/2$.
The target energy is either selected near the mean energy $\epsilon_\text{tar}\approx0$ or at a higher energy $\epsilon_\text{tar}\approx\epsilon_{\rm max}/2$, where $\epsilon_{\rm max}$ is the highest quasiparticle energy corresponding to the eigenstate index $\alpha=V$.
The width of the energy window is $\delta=\left(\epsilon_{\rm max}-\epsilon_{\rm min}\right)/100$, where $\epsilon_{\rm min}$ is the lowest quasiparticle energy corresponding to the eigenstate index $\alpha=1$.
We numerically calculate both $\epsilon_\text{tar}$ and $\delta$ for each disorder realization.

\begin{figure}[t!]
\centering
\includegraphics[width=\columnwidth]{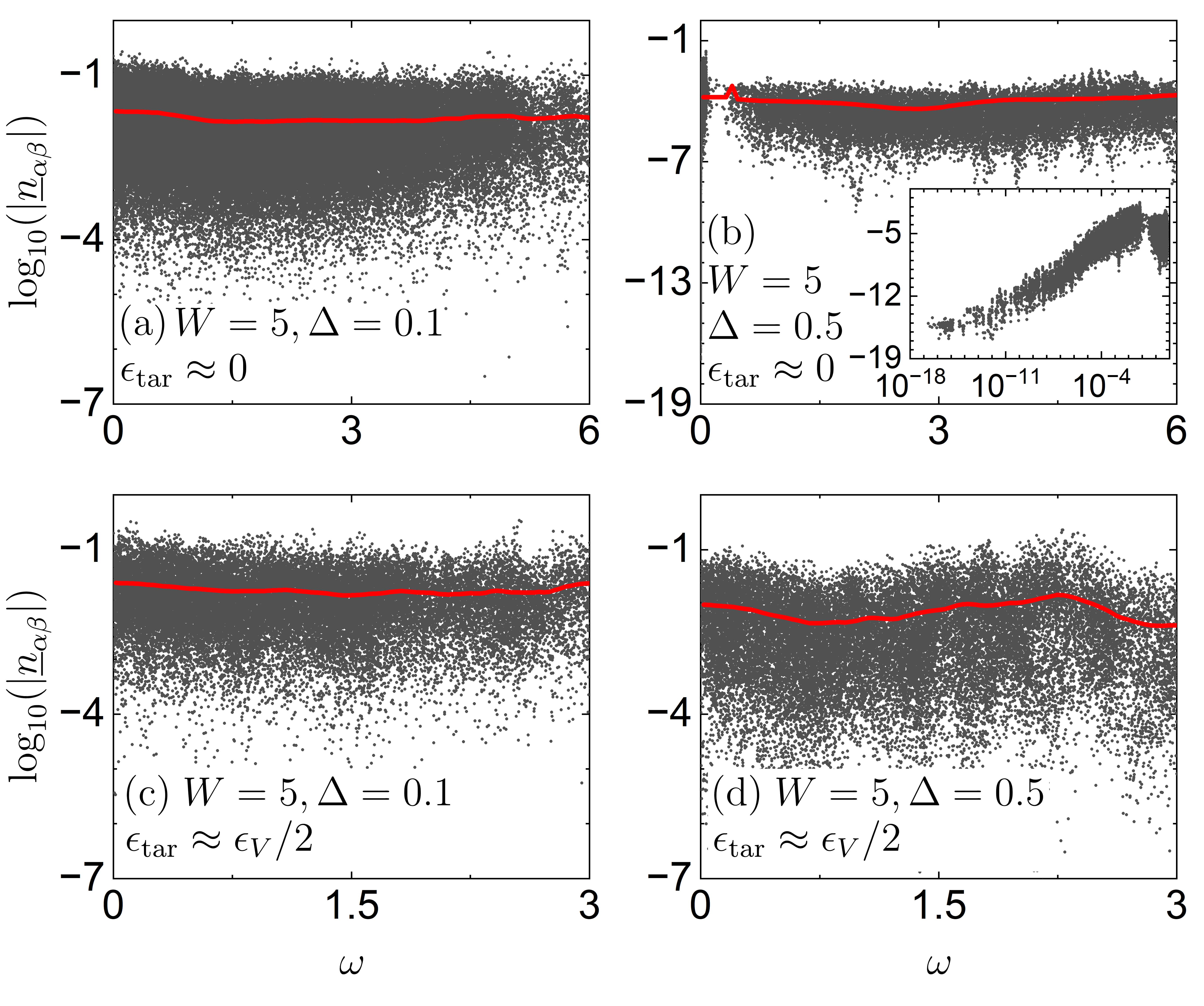}
\caption{Logarithms of absolute values of offdiagonal matrix elements $\log_{10}(|\underline{n}_{\alpha\beta}|)$ as functions of $\omega$, at $V=10^3$ and for $10$ disorder realizations.
The solid red lines denote the coarse-grained offdiagonal matrix elements $\log_{10}(\overline{|\underline{n}_{\alpha\beta}|})$, see text for details.
(a,c) $\Delta=0.1$ and (b,d) $\Delta = 0.5$.
The target energies are (a,b) $\epsilon_\text{tar}\approx0$ and (c,d) $\epsilon_\text{tar}\approx\epsilon_\text{max}/2$.
The inset of (b) displays the same results as the main panel but as a function of $\log_{10}(\omega)$.}
\label{figo1}
\end{figure}

\begin{figure}[t!]
\centering
\includegraphics[width=\columnwidth]{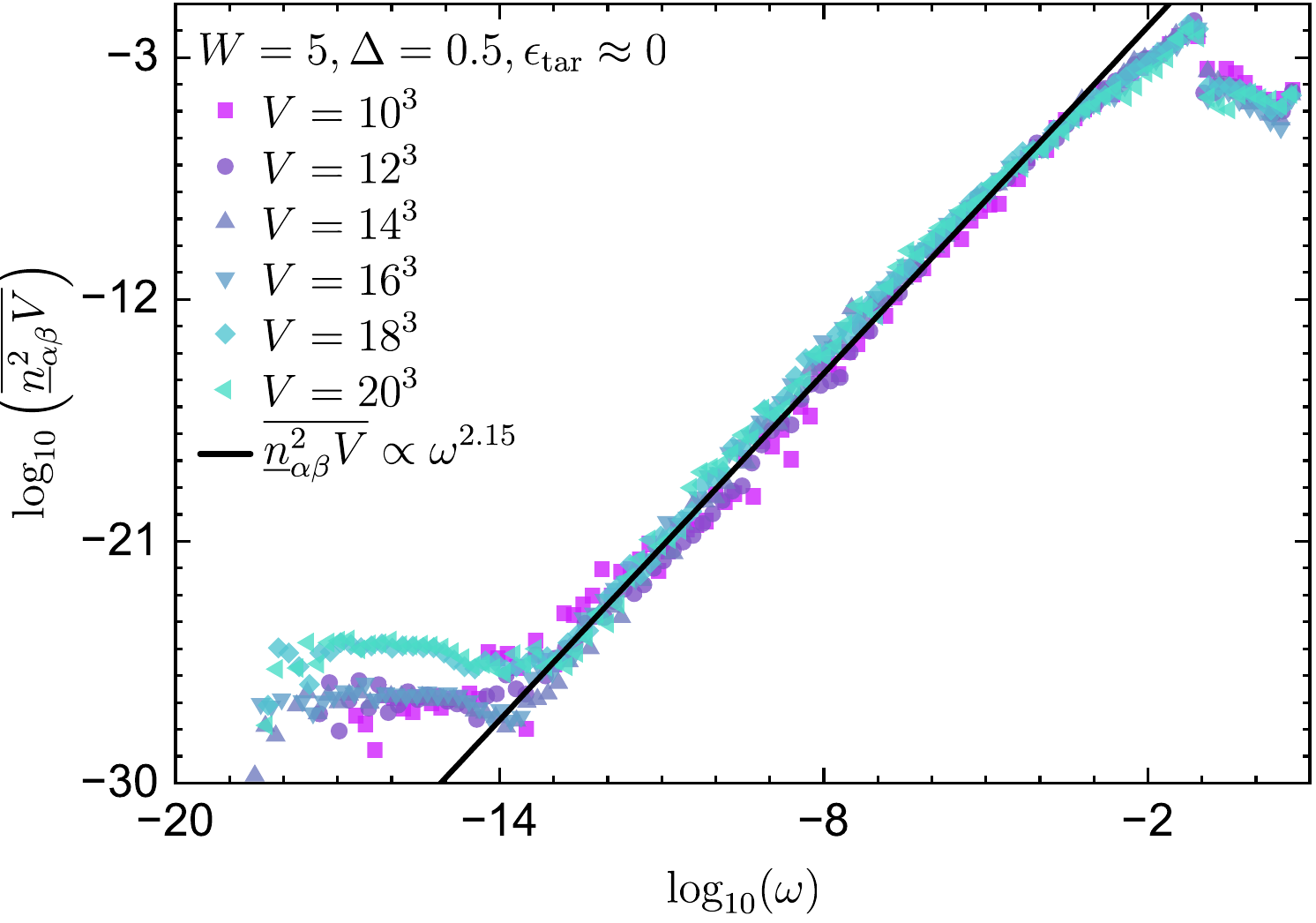}
\caption{Scaled coarse-grained values $\log_{10}\left(\overline{\underline{n}_{\alpha\beta}^2V}\right)$ as functions of $\log_{10}\left(\omega\right)$, at $\Delta=0.5$ and the target energy $\epsilon_\text{tar}\approx 0$.
The results are averaged over $10$ and $2$ disorder realizations for $V\le 18^3$  and $V=20^3$, respectively.
The solid line is the function $\overline{\underline{n}_{\alpha\beta}^2V} \propto \omega^{\eta}$, where $\eta = 2.15$ was obtained using the linear regression.}
\label{figo2}
\end{figure}

In Fig.~\ref{figo1} we plot $\log_{10}(|\underline{n}_{\alpha\beta}|)$ as functions of energy differences $\omega=|\epsilon_\beta-\epsilon_\alpha|$ for a system with $V=10^3$ lattice sites and $10$ disorder realizations.
Solid red curves denote the coarse-grained values $\log_{10}(\overline{|\underline{n}_{\alpha\beta}|})$.
They are obtained by dividing the entire range of $\omega$ into $100$ bins, followed by the averaging of absolute values of offdiagonal matrix elements $|\underline{n}_{\alpha\beta}|$ within each bin and over 10 disorder realizations.

Figures~\ref{figo1}(a) and~\ref{figo1}(c) show results for the pairing strength $\Delta=0.1$ and the target energies $\epsilon_\text{tar}\approx 0$ and $\epsilon_{\rm max}/2$, respectively, while Figs.~\ref{figo1}(b) and~\ref{figo1}(d) show results for the pairing strength $\Delta=0.5$ and the same target energies.
We observe in all cases under considerations that the offdiagonal matrix elements are dense, i.e., there is no finite fraction of offdiagonal matrix elements with values that are lower than or of the order of machine precision.
This property is also observed in single-particle eigenstates of quantum-chaotic quadratic systems with particle number conservation~\cite{lydzba_zhang_21}, as well as in many-body eigenstates of quantum-chaotic interacting~\cite{dalessio_kafri_16} and integrable interacting systems~\cite{leblond_mallayya_19}.

The coarse-grained values $\log_{10}(\overline{|\underline{n}_{\alpha\beta}|})$ in Fig.~\ref{figo1} reveal only a mild dependence on $\omega$, suggesting that the structure function $\mathcal{F}(\overline{\epsilon},\omega)$ from Eq.~(\ref{eqETH}) carries similarities with the random matrix theory (RMT) predictions for which $\mathcal{F}_{\rm RMT}(\omega) \to 1$~\cite{dalessio_kafri_16, schoenle_jansen_21}.
However, as shown in the inset of Fig.~\ref{figo1}(b), the structure function at $\Delta=0.5$ can be nontrivial in the energy range $\epsilon_{\rm tar} \approx 0$ and $\omega\to 0$, in which near zero modes emerge.

It is expected that the offdiagonal matrix elements between zero modes are zero or vanishingly small.
This can be understood by considering a zero mode $\varphi_\beta = (\psi_\alpha+\Gamma\psi_\alpha)/\sqrt{2}$ from Sec.~\ref{sec:general} [below Eq.~(\ref{eq:def_charge})]. For the latter, the wavefunction coefficients $u_{i\alpha}$ and $v_{i\alpha}$ are identical, so that the offdiagonal matrix elements are zero [see Eq.~(\ref{eq:def_nab})].
This agrees with our observation that the offdiagonal matrix elements of near zero modes vanish in the limit $\epsilon_{\rm tar}\to 0$, $\omega\to 0$.

In Fig.~\ref{figo2}, we study the structure function at $\Delta=0.5$ and $\epsilon_{\rm tar} \approx 0$ in the regime of low $\omega$.
To this end, we divide the entire range of $\log_{10}\left(\omega\right)$ into $100$ bins, and average $\underline{n}_{\alpha\beta}^2V$ within each bin, thereby giving rise to the scaled coarse-grained values $\overline{\underline{n}_{\alpha\beta}^2V}$.
For the system sizes under investigation, the scaled results are fairly independent of the system size $V$, and exhibit a polynomial scaling $\overline{\underline{n}_{\alpha\beta}^2V} \propto \omega^{\eta}$, with $\eta\approx 2.15$.
This function interpolates between the vanishing scaled offdiagonal matrix elements $\overline{\underline{n}_{\alpha\beta}^2V} \to 0$ encountered in zero modes (i.e., the limit $\omega\to0$), and $\overline{\underline{n}_{\alpha\beta}^2V} \approx {\rm const}$ at $\omega= O(1)$, which was studied in Fig.~\ref{figo1}.

\subsection{Fluctuations of matrix elements} \label{sec_fluc}

We now perform a quantitative analysis of the fluctuations of matrix elements, focusing on the eigenstate-to-eigenstate fluctuations of the diagonal matrix elements and the variances of both diagonal and offdiagonal matrix elements.

\subsubsection{Eigenstate-to-eigenstate fluctuations}

The eigenstate-to-eigenstate fluctuations determine the differences between diagonal matrix elements in the eigenstates with the nearby single-quasiparticle energies, $\delta \underline{n}_{\alpha}=\underline{n}_{\alpha,\alpha}-\underline{n}_{\alpha-1,\alpha-1}$.
We calculate the average of the absolute values of these differences,
\begin{equation} \label{eq:def_oav}
\delta\underline{n}_\text{av}=||\Lambda||^{-1}\sum_{\ket{\alpha}\in\Lambda} |\delta \underline{n}_{\alpha}| \;,
\end{equation}
as well as the corresponding maximal difference
\begin{equation} \label{eq:def_omax}
\delta\underline{n}_\text{max}=\text{max}_{\ket{\alpha}\in\Lambda} |\delta \underline{n}_{\alpha}|\;,
\end{equation}
where $\Lambda$ is a set comprising either $80\%$ ($||\Lambda||=0.8V$) or $500$ ($||\Lambda||=500$) eigenstates in the middle of the spectrum. We always average $\delta\underline{n}_\text{av}$ and $\delta\underline{n}_\text{av}$ over $100$ disorder realizations to obtain $\langle\langle\delta\underline{n}_\text{av}\rangle\rangle$ and $\langle\langle\delta\underline{n}_\text{max}\rangle\rangle$, respectively.

The eigenstate-to-eigenstate fluctuations from Eqs.~(\ref{eq:def_oav}) and~(\ref{eq:def_omax}) were introduced as indicators of the ETH in many-body eigenstates of interacting Hamiltonians~\cite{Kim_strong2014}.
For single-(quasi)particle eigenstates of quantum-chaotic quadratic Hamiltonians, both indicators are expected to decay to zero as $\propto V^{-\eta}$ with $0<\eta\leq 0.5$~\cite{lydzba_zhang_21}.
A particularly strong indicator of eigenstate thermalization is $\delta\underline{n}_\text{max}$ from Eq.~(\ref{eq:def_omax}), which is expected to decay to zero with increasing $V$ in both quantum-chaotic quadratic models~\cite{lydzba_zhang_21} and quantum-chaotic interacting models~\cite{Mondaini2016,luitz_16,jansen_stolpp_19, leblond_mallayya_19}.

\begin{figure}[t!]
\centering
\includegraphics[width=\columnwidth]{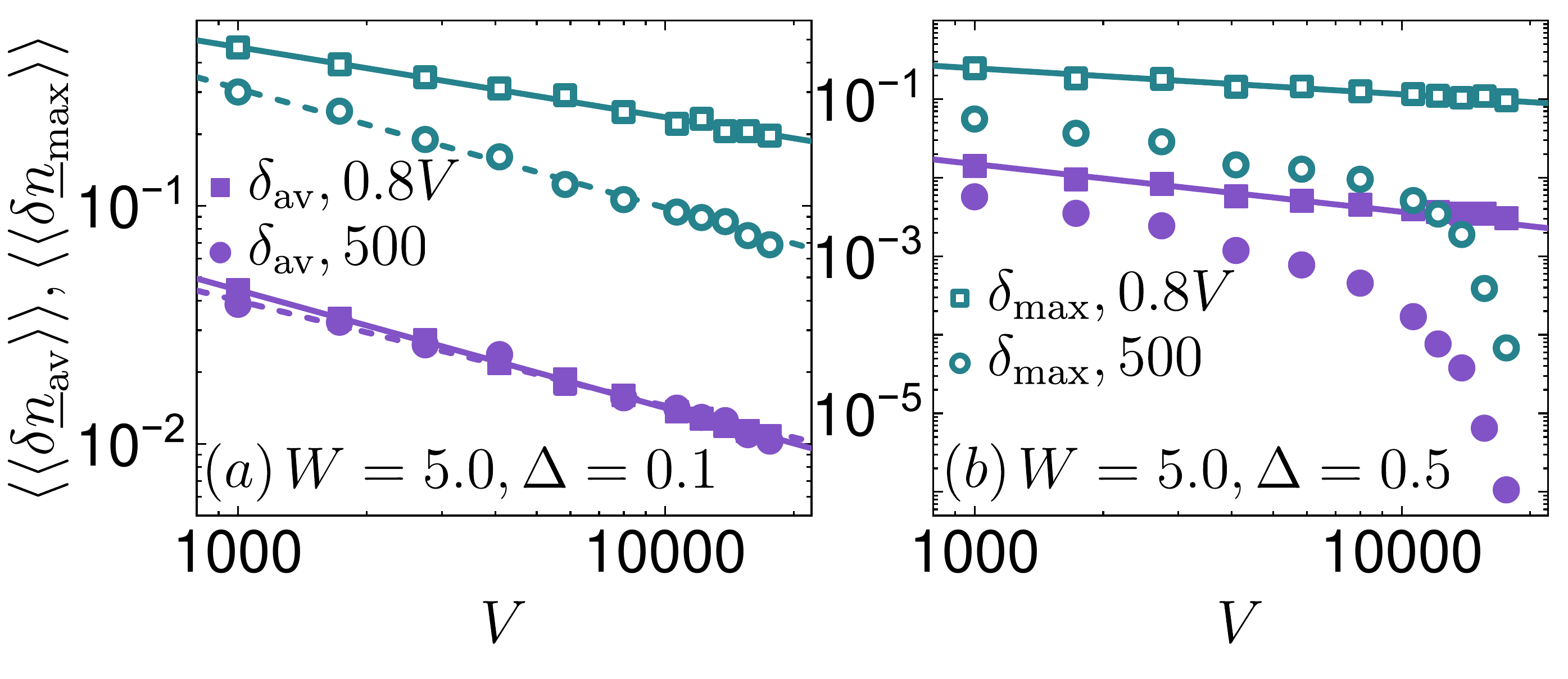}
\caption{Eigenstate-to-eigenstate fluctuations $\langle\langle\delta\underline{n}_\text{av}\rangle\rangle$ (filled symbols) and $\langle\langle\delta\underline{n}_\text{max}\rangle\rangle$ (open symbols) versus the system size $V$, shown at (a) $\Delta=0.1$ and (b) $\Delta = 0.5$.
Squares: averages over $80\%$ of eigenstates.
Circles: averages over $500$ eigenstates in the middle of the spectrum.
Solid and dashed lines are the two-parameter fits of the function $a V^{-\eta}$ to all points. We obtain $\eta\approx 0.5$ ($\eta\approx 0.3$) for $\langle\langle\delta\underline{n}_\text{av}\rangle\rangle$ ($\langle\langle\delta\underline{n}_\text{max}\rangle\rangle$) for arbitrary $\Delta$ and $||\Lambda||=0.8V$. Simultaneously, we obtain $\eta\approx 0.5$ for $\langle\langle\delta\underline{n}_\text{av}\rangle\rangle$ and $\langle\langle\delta\underline{n}_\text{max}\rangle\rangle$ for $\Delta=0.1$ and $||\Lambda||=500$. }
\label{figf1}
\end{figure}

Figures~\ref{figf1}(a) and~\ref{figf1}(b) show the eigenstate-to-eigenstate fluctuations at $\Delta=0.1$ and 0.5, respectively.
They all decay to zero with increasing $V$ and they generally behave as expected for quantum-chaotic quadratic models.
Specifically, $\langle\langle\delta\underline{n}_\text{av}\rangle\rangle \propto V^{-0.5}$ while $\langle\langle\delta\underline{n}_\text{max}\rangle\rangle \propto V^{-\eta}$, with the exponent $\eta\in [0.3,0.5]$.
The observation that $\eta$ in the latter case is slightly smaller than $0.5$ is in agreement with results from other local quantum-chaotic quadratic models, e.g., the 3D Anderson model~\cite{lydzba_zhang_21}.

In Fig.~\ref{figf1} we compare results for $\langle\langle\delta\underline{n}_\text{av}\rangle\rangle$ and $\langle\langle\delta\underline{n}_\text{max}\rangle\rangle$ that are averaged over $80\%$ of eigenstates in the middle of the single-quasiparticle spectrum, with those averaged over $500$ eigenstates in the middle of the spectrum.
While the choice of averaging does not yield any significant differences at $\Delta=0.1$, see Fig.~\ref{figf1}(a), it gives rise to a markedly different scaling with $V$ at $\Delta=0.5$, see Fig.~\ref{figf1}(b).
In particular, when $\Delta=0.5$ and the averages are carried out over 500 eigenstates in the middle of the spectrum, the eigenstate-to-eigenstate fluctuations decay much faster than $\propto V^{-0.5}$.
We comment on this property in more detail in the next section, where a similar scaling is observed for the offdiagonal matrix elements. We can trace it back to the emergence of near zero modes.

\subsubsection{Variances}

\begin{figure}[t!]
\centering
\includegraphics[width=\columnwidth]{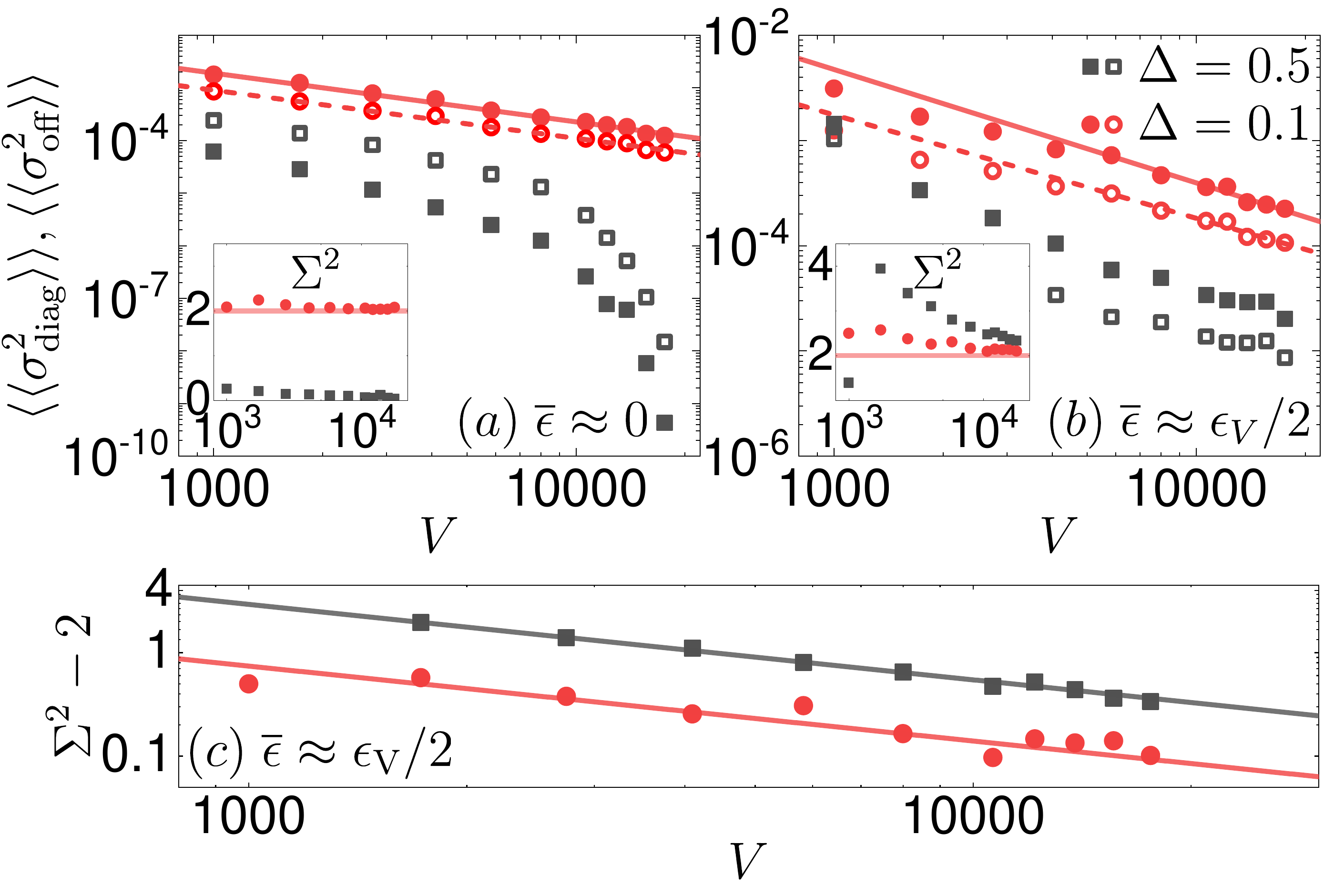}
\caption{Variances $\langle\langle\sigma_\text{diag}^2\rangle\rangle$ (filled symbols) and $\langle\langle\sigma_\text{off}^2\rangle\rangle$ (open symbols) versus the system size $V$, shown at the target energies (a) $\overline{\epsilon}\approx 0$ and (b) $\overline{\epsilon}\approx \epsilon_{\rm max}/2$.
Circles: $\Delta=0.1$, squares: $\Delta=0.5$.
Solid and dashed lines are the two-parameter fits of the function $aV^{-\zeta}$ to the points at $V\ge 18^3$, and we obtain $\zeta\in[0.9,1.1]$.
The insets in (a) and (b) show the ratio $\Sigma^2$ from Eq.~(\ref{eq:def_ratio}), and the horizontal line is $\Sigma_{\rm GOE}^2=2$.
(c) $\Sigma^2-2$ at $\overline{\epsilon}\approx \epsilon_{\rm max}/2$.
Solid lines are two-parameter fits of the function $b V^{-\kappa}$ to the points at $V\ge 10^3$, and we obtain $\kappa\in[0.72,0.73]$.}
\label{figf2}
\end{figure}

Variances of matrix elements are standard indicators of the ETH.
In particular, it has been realized that it is convenient to study variances in narrow energy windows~\cite{mondaini_rigol_17, jansen_stolpp_19}, in which the impact of the structure of matrix elements can be neglected.
We define the variance of diagonal matrix elements as
\begin{equation} \label{eq:def_var1}
\sigma_\text{diag}^2=||\Lambda||^{-1}\sum_{|\alpha\rangle\in\Lambda}\underline{n}_{\alpha\alpha}^2-\Bigg(||\Lambda||^{-1}\sum_{|\alpha\rangle\in\Lambda}\underline{n}_{\alpha\alpha}\Bigg)^2 \;,
\end{equation}
and the variance of offdiagonal matrix elements as
\begin{equation} \label{eq:def_var2}
\sigma_\text{off}^2=||\Lambda'||^{-1}\sum_{\substack{|\alpha\rangle,|\beta\rangle\in\Lambda'\\|\alpha\rangle\neq |\beta\rangle}}\underline{n}_{\alpha\beta}^2-\Bigg(||\Lambda'||^{-1}\sum_{\substack{|\alpha\rangle,|\beta\rangle\in\Lambda'\\|\alpha\rangle\neq |\beta\rangle}}\underline{n}_{\alpha\beta}\Bigg)^2 \;,
\end{equation}
where $\Lambda$ ($\Lambda'$) is a set comprising diagonal (offdiagonal) matrix elements created from $200$ energy eigenstates around the target energy $\overline{\epsilon}$, such that $||\Lambda||=200$ ($||\Lambda'||=39800$). In all calculations, we establish $\sigma_\text{diag}^2$ and $\sigma_\text{off}^2$ for a single disorder realization, which we then average over $100$ disorder realizations. The latter averages are denoted as $\langle\langle\sigma_\text{diag}^2\rangle\rangle$ and $\langle\langle\sigma_\text{off}^2\rangle\rangle$.

We calculate the variances at two target energies, i.e.,  $\overline{\epsilon}\approx 0$ in Fig.~\ref{figf2}(a) and $\overline{\epsilon}\approx \epsilon_{\rm max}/2$ in Fig.~\ref{figf2}(b).
For both cases at $\Delta=0.1$, and for $\overline{\epsilon}\approx \epsilon_{\rm max}/2$ at $\Delta=0.5$, the variances behave as expected for quantum-chaotic quadratic Hamiltonians~\cite{lydzba_zhang_21, ulcakar_vidmar_22, suntajs_prosen_23}.
Specifically, both $\langle\langle\sigma_\text{diag}^2\rangle\rangle$ and $\langle\langle\sigma_\text{off}^2\rangle\rangle$ scale approximately as $1/V$. The least-squares fits of a two-parameter function $aV^{-\zeta}$ are presented in Figs.~\ref{figf2}(a)-\ref{figf2}(b). The exception is $\overline{\epsilon}\approx \epsilon_{\rm max}/2$ at $\Delta=0.5$, for which the least-squares fit is not fully reliable for the accessible $V$. Moreover, the ratio
\begin{equation} \label{eq:def_ratio}
\Sigma^2= \frac{\langle\langle\sigma_\text{diag}^2\rangle\rangle}{\langle\langle\sigma_\text{off}^2\rangle\rangle}\;
\end{equation}
is close to the value $\Sigma^2_{\rm GOE}=2$ predicted by the GOE~\cite{dalessio_kafri_16}, see the insets of Fig.~\ref{figf2}(a) and~\ref{figf2}(b).
The deviation from $\Sigma^2_{\rm GOE}=2$ is expected to be a finite-size effect. Indeed, we show in Fig.~\ref{figf2}(c) that the difference $\Sigma^2-2$ at $\overline{\epsilon}\approx \epsilon_{\rm max}/2$ for both $\Delta=0.1$ and $0.5$ vanishes in the thermodynamic limit as $bV^{-\kappa}$ with $\kappa \approx 0.7$.

Special attention should be devoted to the variances in the middle of the spectrum ($\overline{\epsilon}\approx 0$) at $\Delta=0.5$, for which the results in Fig.~\ref{figf2}(a) show a decay that is much faster than $1/V$.
We argue that such a decay is a consequence of an increasing number of near zero modes that are included in the variances in Eqs.~(\ref{eq:def_var1}) and~(\ref{eq:def_var2}).
Namely, assuming that the number of near zero modes increases as $a_0 V^{1-\zeta}$, with $0<1-\zeta<1$ as suggested by Fig.~\ref{fig6}(a), we develop a two-fluid approximation for the variances in which the squared matrix elements are zero for near zero modes and proportional to $1/V$ otherwise.
This yields (see Appendix~\ref{sec:twofluid} for details)
\begin{equation} \label{eq:def_sigma_twofluid}
    \sigma_{\rm diag}^2 \propto \frac{1}{V} \Big(1-\left(\frac{V}{V^*}\right)^{1-\zeta}\Big)\;,\;\;
    \sigma_{\rm off}^2 \propto \frac{1}{V} \Big(1-\left(\frac{V}{V^*}\right)^{2-2\zeta}\Big) \;,
\end{equation}
where $V^*=(||\Lambda||/a_0)^{1/(1-\zeta)}$ is the number of lattice sites when the number of near zero modes becomes equal to $||\Lambda||$.
For the system sizes under investigation, $V < V^*$, so that the variances are small yet nonzero.
The results in Eq.~(\ref{eq:def_sigma_twofluid}) yield a faster than $1/V$ decay of the variances, which is further discussed in Appendix~\ref{sec:twofluid}.

\section{Quantum quenches and equilibration} \label{sec:quench}

Finally, we turn our attention to the consequences of single-quasiparticle eigenstate thermalization ansatz for the nonequilibrium quantum dynamics.
In a recent work~\cite{lydzba_mierzejewski_23}, it has been demonstrated that this ansatz does not imply thermalization of quadratic systems with a~non-vanishing particle density. Specifically, the infinite-time averages of expectation values of observables disagree (agree) with the predictions of the Gibbs ensemble (the generalized Gibbs ensemble). 
Simultaneously, the equilibration of observables in the many-body states is guaranteed for quadratic Hamiltonians that comply with the single-particle eigenstate thermalization. 
In this section, we show that it is also guaranteed for quadratic Hamiltonians that comply with the single-quasiparticle eigenstate thermalization.

We consider a quantum quench protocol, in which we prepare a system in a pure many-body state $|\Psi_0\rangle$, and evolve it under a Hamiltonian $\hat{H}=\sum_{\alpha=1}^{V}2\epsilon_\alpha \hat{f}_{\alpha}^\dagger\hat{f}_{\alpha}$, for which $|\Psi_0\rangle$ is not an eigenstate.
We focus on one-body observables $\hat{O}= \sum_{i,j=1}^{V} O_{ij}\hat{c}_{i}^\dagger\hat{c}_{j}$ of rank $\mathbf{O}(1)$, namely, on one-body observables that have an $\mathbf{O}(1)$ number of nondegenerate eigenvalues in the single-particle spectrum.
The simplest examples of such observables are the site occupation operator $\hat n_i = \hat c_i^\dagger \hat c_i$ studied in Sec.~\ref{sec:sqETH}, and the quasimomentum occupation operator that in one dimension has the simple form $\hat m_k = \sum_{l,j=1}^V \frac{1}{V} e^{i(l-j)k} \hat c_l^\dagger \hat c_j$.
However, the analysis can be generalized to extensive observables with rank $\mathbf{O}(V)$.

The observables of interest have the following form
\begin{equation}
\label{eqO}
    \hat{O} = \sum_{\alpha,\beta=1}^{V}(O_{\alpha\beta} \hat{f}_{\alpha}^\dagger\hat{f}_{\beta}+A_{\alpha\beta} \hat{f}_{\alpha}^\dagger\hat{f}_{\beta}^\dagger+A_{\alpha\beta}^{*} \hat{f}_{\beta}\hat{f}_{\alpha}),
\end{equation}
where $O_{\alpha\beta}=\langle \alpha | \hat{O} | \beta\rangle=\sum_{i,j=1}^{V} O_{ij}(u_{i\alpha}u_{j\beta}-v_{i\alpha}v_{j\beta})$ is hermitian, while $A_{\alpha\beta}=\langle\alpha|\hat{O}\Gamma|\beta\rangle=\sum_{i,j=1}^{V} O_{ij}u_{i\alpha}v_{j\beta}$ is antisymmetric. We can rewrite $\hat{O}$ in the Nambu representation
\begin{equation} \label{eq:Omatrix}
\hat{O}=\hat{F}^\dagger 
    \begin{bmatrix}
    \frac{1}{2}{O} & A\\
    -A^* & -\frac{1}{2}O^{*}
    \end{bmatrix}
\hat{F}
= \hat{F}^\dagger \mathcal{O} \hat{F},
\end{equation}
where $\hat{F}=\left[\hat{f}_{1}\;...\;\hat{f}_{V}\;\hat{f}_{1}^\dagger\;...\;\hat{f}_{V}^\dagger\right]^\text{T}$ is a $2V\times 1$ vector introduced in Sec.~\ref{sec:general} and the matrix $\cal O$ should be distinguished from the smooth function ${\cal O}(\bar \epsilon)$ in Eq.~(\ref{eqETH}).

Note that the matrix $O$ from Eq.~(\ref{eq:Omatrix}) can be understood as being composed of matrix elements of $\hat{O}$ between energy eigenstates from the same symmetry sector, while the matrix $A$ as being composed of matrix elements of $\hat{O}$ between energy eigenstates from different symmetry sectors. As argued in Ref.~\cite{LeBlond_2020}, the behaviour (e.g., the scaling of the variance) of offdiagonal matrix elements that connect energy eigenstates from the same versus different symmetry sectors is qualitatively similar.

We are interested in the time evolution, so we express $\hat{O}$ in the Heisenberg representation
\begin{align}
     &\hat{O}(t)=e^{i\hat{H}t}\hat{O}e^{-i\hat{H}t} =
     \sum_{\alpha,\beta=1}^{V} O_{\alpha\beta}e^{2i(\epsilon_\alpha-\epsilon_\beta)t}\hat{f}_{\alpha}^\dagger \hat{f}_\beta\\
     & + \sum_{\alpha,\beta=1}^{V} A_{\alpha\beta}e^{2i(\epsilon_\alpha+\epsilon_\beta)t}\hat{f}_{\alpha}^\dagger \hat{f}_\beta^\dagger
     + \sum_{\alpha,\beta=1}^{V} A_{\alpha\beta}^{*}e^{-2i(\epsilon_\alpha+\epsilon_\beta)t}\hat{f}_{\beta} \hat{f}_\alpha, \nonumber
\end{align}
and calculate its expectation value in the initial state~$|\Psi_0\rangle$
\begin{align}
    & \langle\hat{O}(t)\rangle = 
    \sum_{\alpha,\beta=1}^{V} O_{\alpha\beta}e^{2i(\epsilon_\alpha-\epsilon_\beta)t} R_{\alpha\beta}\\
    & + \sum_{\alpha,\beta=1}^{V} A_{\alpha\beta}e^{2i(\epsilon_\alpha+\epsilon_\beta)t} M_{\alpha\beta}
    + \sum_{\alpha,\beta=1}^{V} A_{\alpha\beta}^{*}e^{-2i(\epsilon_\alpha+\epsilon_\beta)t} M_{\alpha\beta}^{*}, \nonumber
\end{align}
where $R_{\alpha\beta}=\langle\Psi_0|\hat{f}_{\alpha}^\dagger \hat{f}_\beta|\Psi_0\rangle$ and $M_{\alpha\beta}=\langle\Psi_0|\hat{f}_{\alpha}^\dagger \hat{f}_\beta^\dagger|\Psi_0\rangle$. These matrices are parts of the generalized one-body correlation matrix
\begin{equation}
    \mathcal{\rho}=
    \begin{bmatrix}
    R & M\\
    -M^* & -R^{*}
    \end{bmatrix}\;,
\end{equation}
which eigenvalues for fermions belong to the interval $[0,1]$, so that $\text{Tr}[{\mathcal{\rho}^2}]\le\text{Tr}[{\mathcal{\rho}}]\le 2V$~\cite{Chung_2001,Peschel_2003,Cheong_2004,Peschel_2009,Vidmar_2017,Hackl_2021}.

\begin{figure}[t!]
\centering
\includegraphics[width=\columnwidth]{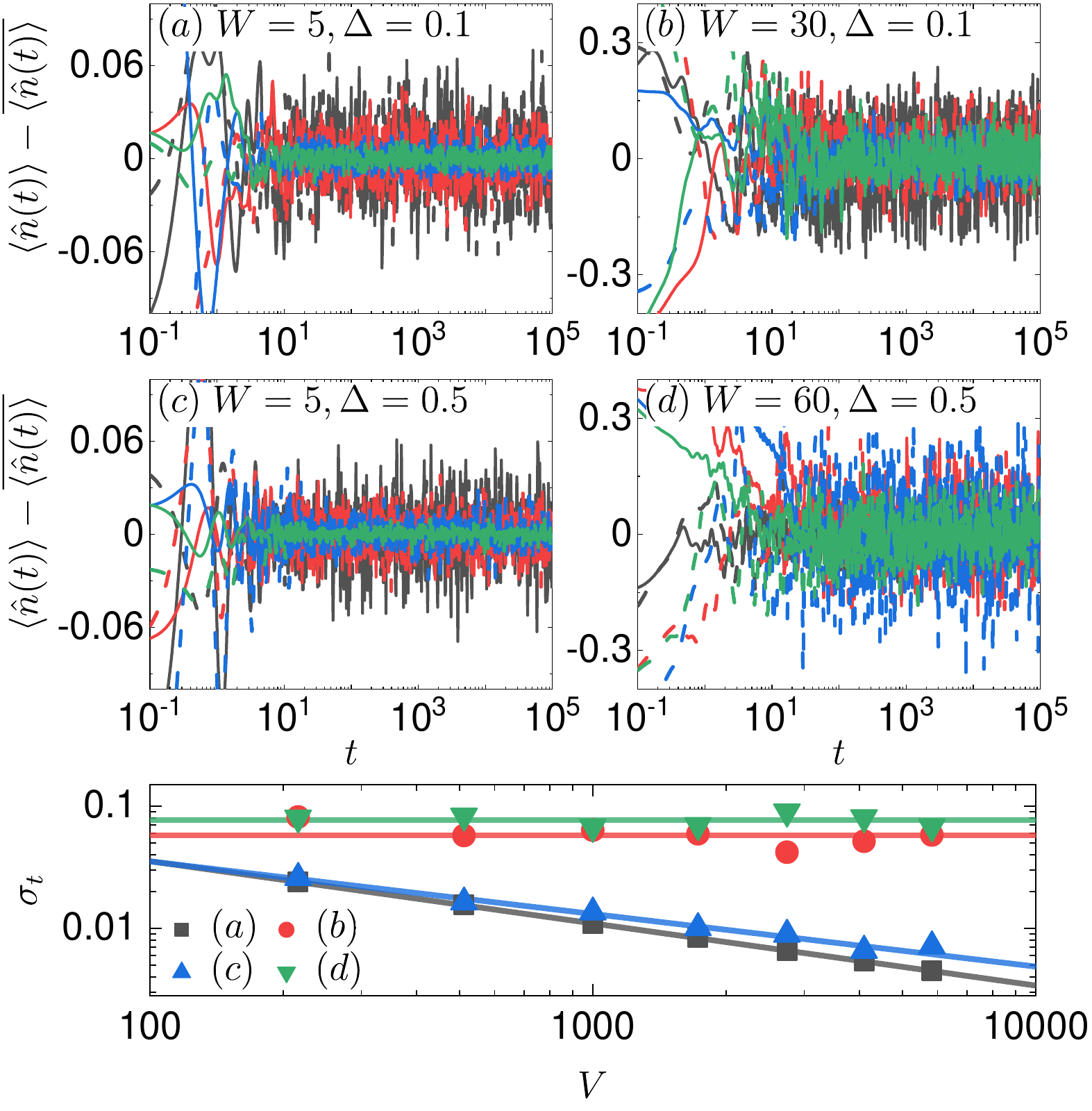}
\caption{(a-d) Time evolution of the site occupation $\hat{n}$ in the quantum quench from the initial disorder $\tilde{W}=W+5$ to the disorder $W$ (and a different disorder realization).
(a) $W=5$ and $\Delta=0.1$, (b) $W=30$ and $\Delta=0.1$, (c) $W=5$ and $\Delta=0.5$, (d) $W=60$ and $\Delta=0.1$.
Numerical results at $V = 6^3, 8^3, 14^3$, and $18^3$ are marked with black, red, blue, and green, respectively.
We show results for two (solid and dashed) quench realizations for each $V$.
(e) Temporal fluctuations $\sigma_t$ from Eq.~(\ref{eq:def_sigmat}), calculated within the time interval $t\in[10^2, 10^5]$ and averaged over $20$ quench realizations.
The solid lines show the two-parameter fits of the function $c V^{-\zeta}$. We get $\zeta\in[0.4, 0.5]$ for (a) and (c).}
\label{figt1}
\end{figure}

The equilibration of an observable means that the temporal fluctuations of its expectation value about the infinite-time average vanish in the thermodynamic limit. The temporal fluctuations can be characterized by the variance
\begin{equation} \label{eq:def_sigmat}
    \sigma_t^2 = \overline{\langle\hat{O}(t)\rangle^2}-\overline{\langle\hat{O}(t)\rangle}^2,
\end{equation}
where $\overline{\langle\hat{O}(t)\rangle}=\text{lim}_{\tau\rightarrow\infty}\int_{0}^{\tau}\langle\hat{O}(t)\rangle dt$ is the infinite time average. 
We calculate the time average as
\begin{align} \label{eq:Ot_avrage}
    &\overline{\langle\hat{O}(t)\rangle} =\sum_{\alpha,\beta=1}^{V} O_{\alpha\beta} R_{\alpha\beta} \overline{e^{2i(\epsilon_\alpha-\epsilon_\beta)t}}\\
    & + \sum_{\alpha,\beta=1}^{V} A_{\alpha\beta} M_{\alpha\beta} \overline{e^{2i(\epsilon_\alpha+\epsilon_\beta)t}}
    + \sum_{\alpha,\beta=1}^{V} A_{\alpha\beta}^{*} M_{\alpha\beta}^{*} \overline{e^{-2i(\epsilon_\alpha+\epsilon_\beta)t}}. \nonumber
\end{align}
In quantum-chaotic quadratic models, there are no degeneracies in the single-quasiparticle energy spectrum.
Therefore, the first term on the r.h.s.~of Eq.~(\ref{eq:Ot_avrage}) is nonvanishing if and only if $\alpha=\beta$, while the last two terms vanish. The infinite time average hence simplifies to $\overline{\langle\hat{O}(t)\rangle}=\sum_{\alpha=1}^{V}O_{\alpha\alpha}R_{\alpha\alpha}$. Using similar arguments and the assumption of no gap degeneracies in the single-quasiparticle energy spectrum, we arrive at
\begin{equation}
\begin{split}
    \overline{\langle\hat{O}(t)^2\rangle} & =\overline{\langle\hat{O}(t)\rangle}^2+\sum_{\alpha\neq\beta}|O_{\alpha\beta}|^2|R_{\alpha\beta}|^2\\
    & + 4\sum_{\alpha,\beta=1}^{V}|A_{\alpha\beta}|^2|M_{\alpha\beta}|^2,
\end{split}
\end{equation}
so that the variance is given by
\begin{equation}
    \sigma_t^2=\sum_{\alpha\neq\beta}|O_{\alpha\beta}|^2|R_{\alpha\beta}|^2 + 4\sum_{\alpha,\beta=1}^{V}|A_{\alpha\beta}|^2|M_{\alpha\beta}|^2.
\end{equation}
To determine the upper bound for the temporal fluctuations, we note that $\forall_{\alpha\neq\beta}(|\frac{1}{2}O_{\alpha\beta}|^2\le|\text{max}(\mathcal{O}_{\tilde{\alpha}\tilde{\beta}})|^2)$ and $\forall_{\alpha,\beta}(|A_{\alpha\beta}|^2\le|\text{max}(\mathcal{O}_{\tilde{\alpha}\tilde{\beta}})|^2)$, where $\tilde{\alpha},\tilde{\beta}\in\{1,2,...,2V\}$ run over the Bogoliubov-de Gennes basis, while $\text{max}(\mathcal{O}_{\tilde{\alpha}\tilde{\beta}})$ is the maximal offdiagonal matrix element of $\mathcal{O}$ from Eq.~(\ref{eq:Omatrix}). We can now write
\begin{equation} \label{eq:sigmat_short}
    \sigma_t^2 \le 4|\text{max}(\mathcal{O}_{\tilde{\alpha}\tilde{\beta}})|^2 \sum_{\alpha,\beta}\left(|R_{\alpha\beta}|^2+|M_{\alpha\beta}|^2\right)\;.
\end{equation}
Since the double sum in Eq.~(\ref{eq:sigmat_short}) simplifies to $\sum_{\alpha,\beta}(|R_{\alpha\beta}|^2+|M_{\alpha\beta}|^2)=\text{Tr}(R^2-MM^{*})=\frac{1}{2}\text{Tr}(\mathcal{\rho}^2)$, we obtain
\begin{equation} \label{eq:sigmat_bound}
    \sigma_t^2 \le 2|\text{max}(\mathcal{O}_{\tilde{\alpha}\tilde{\beta}})|^2\text{Tr}(\mathcal{\rho}^2)\le 4 |\text{max}(\mathcal{O}_{\tilde{\alpha}\tilde{\beta}})|^2 V\;.
\end{equation}
Since the relation between one-body observables and their counterparts normalized in the single-quasiparticle sector is $\hat{O}\simeq \underline{\hat{O}}\sqrt{V}$ (see Eq.~(\ref{eqnorm}) and Ref.~\cite{lydzba_zhang_21}), the single-quasiparticle eigenstate thermalization in quantum-chaotic quadratic Hamiltonians results in $|\text{max}(\mathcal{O}_{\tilde{\alpha}\tilde{\beta}})|^2 V\simeq |\text{max}(\mathcal{\underline{O}}_{\tilde{\alpha}\tilde{\beta}})|^2\propto 1/V$. Hence, the equilibration of
these one-body observables is guaranteed in the thermodynamic limit.

In Fig.~\ref{figt1}, we numerically test equilibration of a site occupation operator~$\hat{n}$, for which matrix elements in single-quasiparticle eigenstates were studied in Sec.~\ref{sec:sqETH}.
In the quantum quench setup, the system is prepared in the many-quasiparticle ground state at disorder $\tilde{W}=W+5$.
Specifically, we construct a state with $N=V/2$ quasiparticles with the lowest positive energies~$\epsilon_\alpha$ on top of the many-particle ground state (i.e., the Bogoliubov vacuum).
The state is evolved with a Hamiltonian at disorder $W$ and a different disorder realization.
Figures~\ref{figt1}(a)-\ref{figt1}(d) show the time evolution of $\langle\hat{n}(t)\rangle-\overline{\langle\hat{n}(t)\rangle}$, where $\overline{\langle\hat{n}(t)\rangle}$ is the long-time average calculated within the time interval $t\in[10^2, 10^5]$.
The temporal fluctuations $\sigma_t$ as functions of $V$ are shown in Fig.~\ref{figt1}(e). 

For the quenches at $W=5$, see Fig.~\ref{figt1}(a) [$\Delta=0.1$] and Fig.~\ref{figt1}(c) [$\Delta=0.5$], the temporal fluctuations decrease with increasing system size, and a scaling $\sigma_t\propto V^{-\zeta}$ with $\zeta\approx0.5$ is observed in Fig.~\ref{figt1}(e).
For this choice of parameters, the Hamiltonian is quantum-chaotic quadratic and the single-quasiparticle eigenstate thermalization was demonstrated in Sec.~\ref{sec:sqETH}.
Then, equilibration of the observable is guaranteed by Eq.~(\ref{eq:sigmat_bound}).

The situation is different at a large disorder when the system exhibits signatures of localization, as demonstrated in Fig.~\ref{fig1} by the IPR analysis.
Examples are shown in Fig.~\ref{figt1}(b) [$W=30$, $\Delta=0.1$] and Fig.~\ref{figt1}(d) [$W=60$, $\Delta=0.5$].
In these cases, the temporal fluctuations do not decrease with increasing system size and a scaling $\sigma_t\propto V^{-\zeta}$ with $\zeta\approx0$ is observed in Fig.~\ref{figt1}(e).

\section{Conclusions} \label{sec:conclusions}

We studied the wavefunction properties and the observable matrix elements in single-quasiparticle eigenstates of a quantum-chaotic quadratic Hamiltonian without the particle number conservation.
Our main goal was to extend the concept of single-particle quantum chaos to general quadratic Hamiltonians.
In particular, we introduced the notion of single-quasiparticle eigenstate thermalization, which is the ansatz for the matrix elements of observables in single-quasiparticle eigenstates.

We also argued about an important consequence of single-quasiparticle eigenstate thermalization: it guarantees equilibration of observables in the long-time dynamics after quantum quenches at arbitrary quasiparticle numbers.
The proof of equilibration caries analogies with the recent proof of equilibration in quantum-chaotic quadratic Hamiltonians with the particle number conservation~\cite{lydzba_mierzejewski_23}.
Although we only considered one-body observables, we expect the proof to be valid for multi-body observables, as in the Hamiltonians with the particle number conservation~\cite{lydzba_mierzejewski_23}.

Beside these generic features, the Hamiltonian under investigation may also exhibit a considerable number of zero and near zero modes.
When present, they manifest themselves as a sharp peak in the single-quasiparticle density of states and may violate the single-quasiparticle eigenstate thermalization ansatz.
However, their relative number appears to vanish in the thermodynamic limit, and in this sense they carry similarities with the zero modes recently observed in quantum-chaotic tight-binding billiards~\cite{ulcakar_vidmar_22}.
The role of zero modes in nonequilibrium quantum dynamics after quantum quenches is an interesting topic that deserves more attention in future works.

\section*{Acknowledgements}

We acknowledge discussions with M. Mierzejewski, and the support from the Slovenian Research Agency (ARRS), Research core funding Grants No.~P1-0044 and No.~N1-0273 (L.V.). Numerical studies in this
work have been partially carried out using resources provided by
the Wroclaw Centre for Networking and Supercomputing \cite{wcss}, Grant No. 579 (P. T., P. Ł.).

\appendix

\section{Scaling of variances of matrix elements in the middle of the spectrum} \label{sec:twofluid}

In Fig.~\ref{figf2}(a) we observed a rapid decrease of variances with $V$ at $W=5$ and $\Delta=0.5$.
This occurs in the middle of the spectrum ($\bar\epsilon \approx 0$), in which near zero and zero modes are included in the calculations of the variances in Eqs.~(\ref{eq:def_var1}) and~(\ref{eq:def_var2}).
As argued in Sec.~\ref{sec:structure}, the zero modes exhibit diagonal matrix elements that are vanishingly small and independent of the eigenstate index $\alpha$, and offdiagonal matrix elements that are zero.

To understand the rapid decrease of variances with $V$, we develop a two-fluid model in which the squares of matrix elements are either exactly zero (contribution from zero modes) or they scale as $1/V$ (contributions from other states).
We consider the following ansatz, inspired by Fig.~\ref{fig6}(a), for the number $\cal N$ of zero modes: ${\cal N} = a_0 V^{1-\zeta}$, where $a_0$ is a constant and $0<1-\zeta<1$.
Since the total number of diagonal matrix elements included in the variance is $||\Lambda||$, the number of nonzero matrix elements is then $||\tilde{\Lambda}|| = ||\Lambda|| - {\cal N}$.
The system size $V^* = (||\Lambda||/a_0)^{1/(1-\zeta)}$ at which the variance is determined by the zero modes only is given by the condition $||\tilde{\Lambda}||=0$.
In our calculations, we consider $V < V^*$.

The nonzero contribution to the variance of the diagonal matrix elements is then
\begin{equation}
    \sigma_{\rm diag}^2 = \frac{1}{||\Lambda||}\frac{||\tilde{\Lambda}||}{V} \left( \frac{1}{||\tilde{\Lambda}||} \sum_{\alpha\in\Lambda'} V \underline{n}_{\alpha\alpha}^2 \right) \;.
\end{equation}
If the system supports the single-quasiparticle eigenstate thermalization, the term in the parenthesis is a constant. We write it symbolically as~$c_{\rm diag}$.
The variance is then simplified to
\begin{equation}
\label{eq_vardiag}
    \sigma_{\rm diag}^2 = \frac{1}{V}\, c_{\rm diag} \Bigg(1-\left(\frac{V}{V^*}\right)^{1-\zeta}\Bigg) \;.
\end{equation}
In the case of the offdiagonal matrix elements, the total number of matrix elements included in the variance is roughly $||\Lambda||^2$ and the number of nonzero matrix elements is then roughly $||\tilde{\Lambda}||^2 = ||\Lambda||^2 - {\cal N}^2$.
The  analysis similar to the one carried out for the diagonal matrix elements yields
\begin{equation}
    \sigma_{\rm off}^2 = \frac{1}{||\Lambda||^2}\frac{||\tilde{\Lambda}||^2}{V} \left( \frac{1}{||\tilde{\Lambda}||^2} \sum_{\alpha,\beta\in\Lambda'} V \underline{n}_{\alpha\beta}^2 \right) \;.
\end{equation}
Again, if the system supports the single-quasiparticle eigenstate thermalization, the term in the parenthesis is a constant. We write it symbolically as~$c_{\rm off}$. 
The variance then takes the following form
\begin{equation}
\label{eq_varoff}
    \sigma_{\rm off}^2 = \frac{1}{V}\, c_{\rm off} \Bigg(1-\left(\frac{V}{V^*}\right)^{2-2\zeta}\Bigg) \;.
\end{equation}

\begin{figure}[t!]
\centering
\includegraphics[width=\columnwidth]{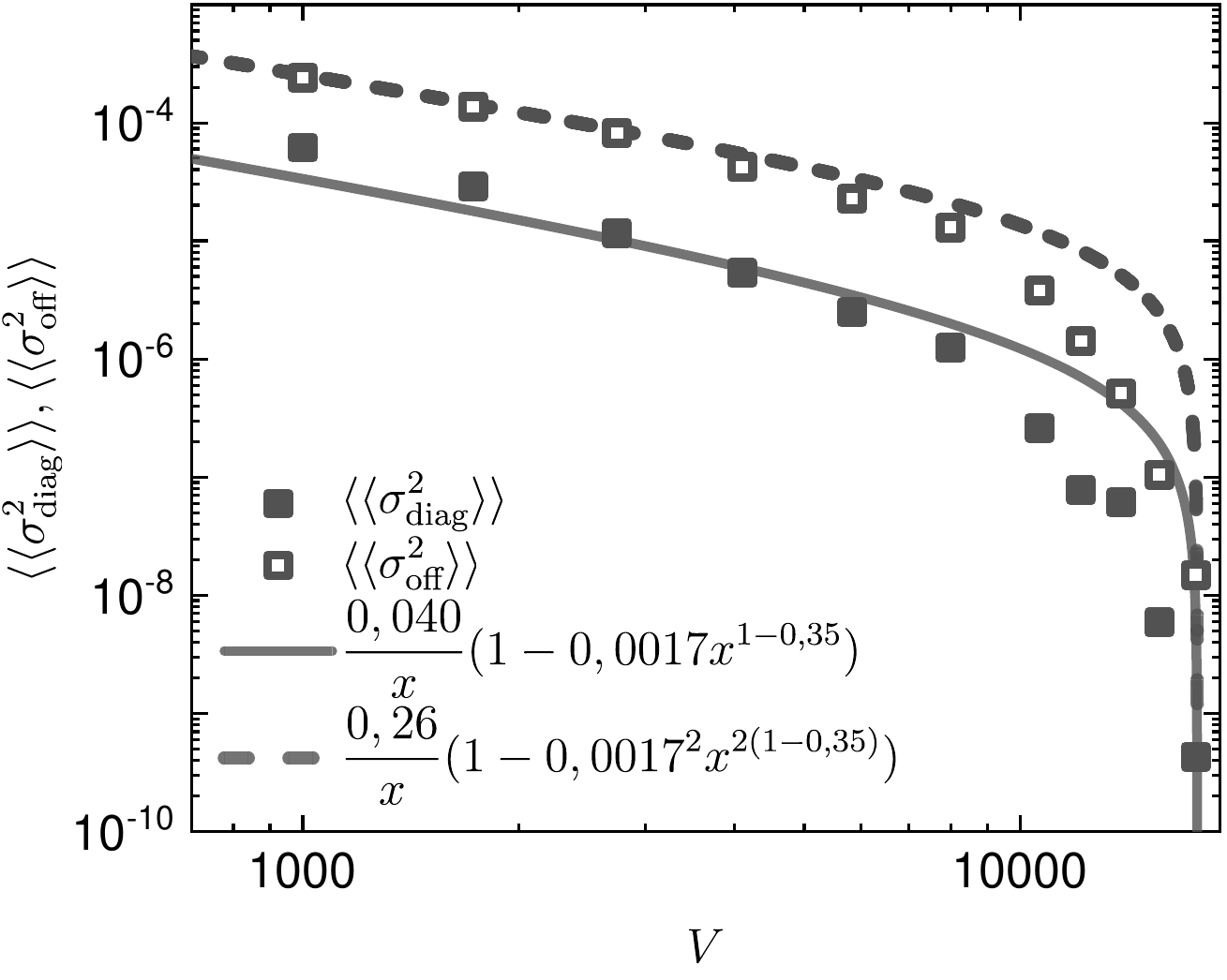}
\caption{Variances $\langle\langle\sigma_\text{diag}^2\rangle\rangle$ (filled symbols) and $\langle\langle\sigma_\text{off}^2\rangle\rangle$ (open symbols) versus the system size $V$ in the middle of the spectrum $\overline{\epsilon}\approx 0$. Solid and dashed lines are the least-squares fits of the analytical expressions from Eqs.~(\ref{eq_vardiag}) and (\ref{eq_varoff}) to all points. We obtain $c_{\rm diag}\approx 0.040$, $c_{\rm off}\approx 0.26$, $\zeta\approx 0.35$ and $V^{*}\approx 18233$.
}
\label{figv1}
\end{figure}

We demonstrate in Fig.~\ref{figv1} that the finite-size scalings of $\sigma^2_\text{diag}$ and $\sigma^2_\text{off}$, which have been established in Eqs.~(\ref{eq_vardiag}) and (\ref{eq_varoff}), qualitatively agree with the numerical simulations, which have been presented in Fig.~\ref{figf2}(a). The variances are proportional to $1/V$ when $V\ll V^*$, and they begin to decrease much faster when $V\rightarrow V^*$. However, there are some quantitative differences. They most likely follow from the fact that the number of zero modes is not exactly polynomial in the system size $V$ (see Fig.~\ref{fig6}), and that the matrix elements of near zero modes have an additional structure in the energy difference $\omega$ (see Figs.~\ref{figd1} and \ref{figo1}) that we neglected. It is also somewhat surprising that $c_\text{diag}<c_\text{off}$. Further improvements of the two-fluid model are out of the scope of this manuscript.

\section{Distributions of matrix elements}

We complement the results in Sec.~\ref{sec:sqETH} by studying the distributions of matrix elements of $\underline{\hat n}$.
It was recently shown that one-body observables in single-particle eigenstates of quantum-chaotic quadratic Hamiltonians may not be Gaussian~\cite{lydzba_zhang_21} (see also~\cite{khaymovich_haque_19, wittmann_castro_22}), and we here expect non-Gaussian distributions are well.
As a simple analytical prediction we consider the approximation in which the coefficients $u_{i\alpha}$ and $v_{i\alpha}$ are 
random variables drawn from the normal distribution with zero mean and variance $\sigma^2=\frac{1}{2V}$,
\begin{equation}
    P_{u}(x)=P_{v}(x)=\frac{1}{\sqrt{2\pi\sigma^2}}\exp(-x^2/{2\sigma^2})\;.
\end{equation}
This approximation, also referred to as the RMT approximation, gave rise to very accurate predictions for the distributions of matrix elements in the 3D Anderson model~\cite{lydzba_zhang_21} and tight-binding billiards~\cite{ulcakar_vidmar_22}.

The RMT approximation simplifies the following sums of coefficients:
$\sum_{\alpha=1}^{V} u_{i\alpha}^2 \to \frac{1}{2}$, $\sum_{\alpha=1}^{V} v_{i\alpha}^2 \to \frac{1}{2}$, and $\sum_{\alpha=1}^{V}u_{i\alpha}v_{i\alpha} \to 0$, which in turn simplify the expressions for the norm and the trace,
\begin{equation}
    \text{N}\to\frac{1}{4V}+\frac{V-1}{4V}+\frac{2}{4V}-\frac{1}{4}=\frac{1}{2V}\,,\;\; \text{T}\to\frac{1}{2}\;.
\end{equation}
The diagonal matrix elements of the normalized operator $\underline{\hat n}$ from Eq.~(\ref{eq:def_n_normalized}) are then
\begin{equation} \label{eq:def_naa_random}
    \underline{n}_{\alpha\alpha}=\sqrt{2V}\left(u_{i\alpha}^2-v_{i\alpha}^2\right)
\end{equation}
and the offdiagonal matrix elements are
\begin{equation} \label{eq:def_nab_random}
    \underline{n}_{\alpha\beta}=\sqrt{2V}\left(u_{i\alpha}u_{i\beta}-v_{i\alpha}v_{i\beta}\right)\,.
\end{equation}

Using these expressions, and some algebra of random variables that is reported in Appendices~\ref{sec:distribution_diag} and~\ref{sec:distribution_off}, one obtains predictions for the distributions of matrix elements.
The distribution of diagonal matrix elements $\underline{n}_{\alpha\alpha}$ is
\begin{equation}
\label{PDF_1}
    P_{\underline{n}_{\alpha\alpha}}(x)=\frac{1}{\pi}\sqrt{\frac{V}{2}}\text{K}_0\left(\sqrt{\frac{V}{2}}|x|\right),
\end{equation}
where $\text{K}_0$ stands for the modified Bessel function of the second kind. 
The distribution of offdiagonal matrix elements $\underline{n}_{\alpha\beta}$ is
\begin{equation}
\label{PDF_2}
     P_{\underline{n}_{\alpha\beta}}(x) = \sqrt{\frac{V}{2}}\exp\left(-\sqrt{2V} |x| \right)\;,
\end{equation}
see Appendices~\ref{sec:distribution_diag} and~\ref{sec:distribution_off} for the details of the derivation.

\begin{figure}[t!]
\centering
\includegraphics[width=\columnwidth]{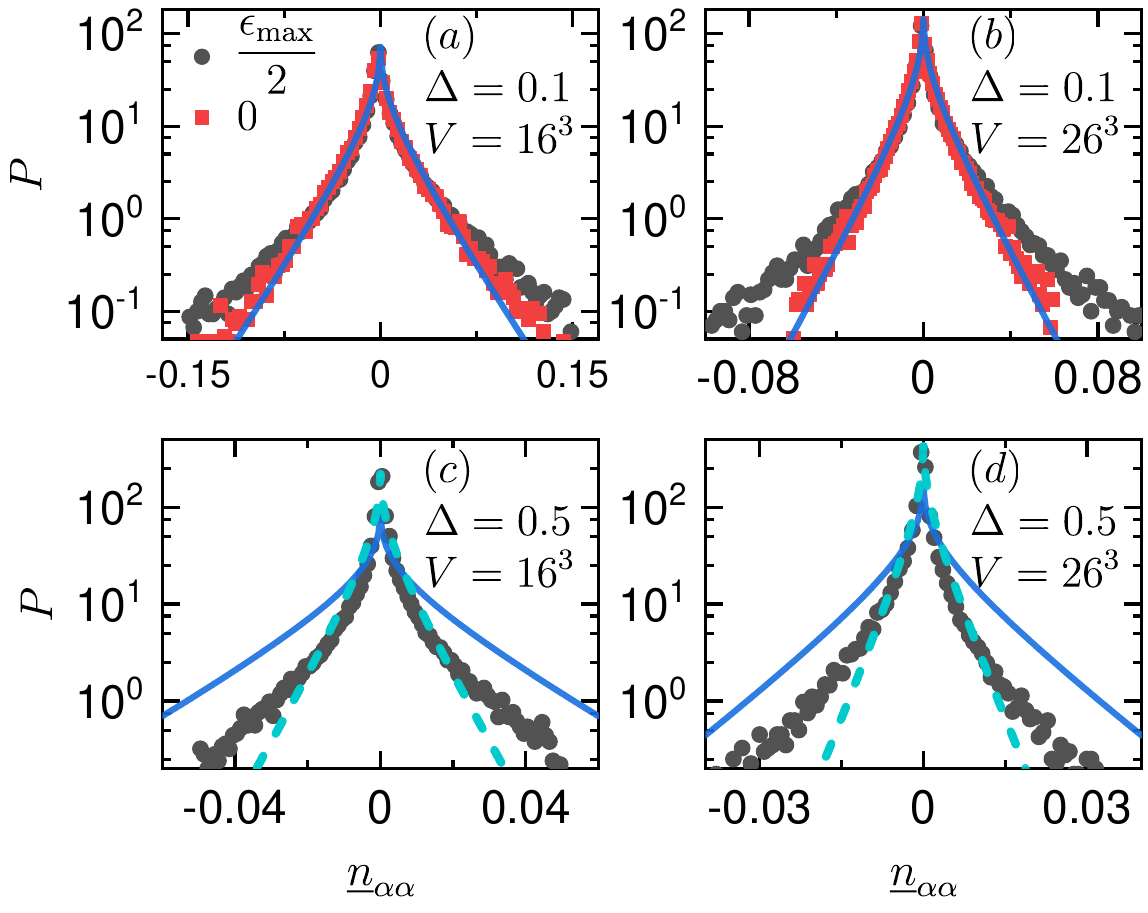}
\caption{Distributions of the diagonal matrix elements $\underline{n}_{\alpha\alpha}$. 
(a) $\Delta=0.1$ and $V=16^3$, (b) $\Delta=0.1$ and $V=26^3$, (c) $\Delta=0.5$ and $V=16^3$, (d) $\Delta=0.5$ and $V=26^3$.
Grey circles (red squares) mark numerical results for $200$ eigenstates near $\overline{\epsilon}\approx \epsilon_{\rm max}/2$ ($\overline{\epsilon}\approx 0$).
Solid lines are results from Eq.~(\ref{PDF_1}), dashed lines are the rescaled results using Eq.~(\ref{eq:dist_rescaled1}).
}
\label{figdi1}
\end{figure}

\begin{figure}[t!]
\centering
\includegraphics[width=\columnwidth]{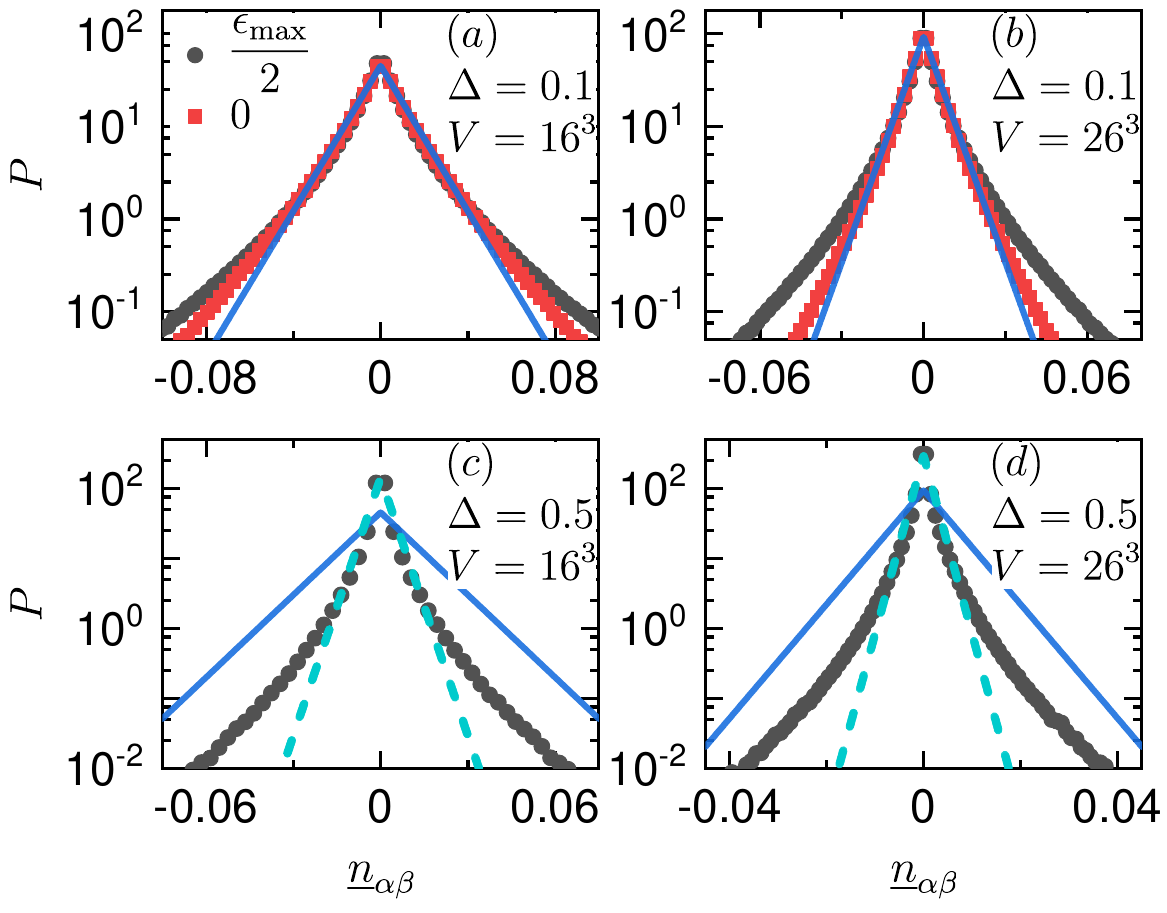}
\caption{Distributions of the offdiagonal matrix elements $\underline{n}_{\alpha\beta}$. 
(a) $\Delta=0.1$ and $V=16^3$, (b) $\Delta=0.1$ and $V=26^3$, (c) $\Delta=0.5$ and $V=16^3$, (d) $\Delta=0.5$ and $V=26^3$.
Grey circles (red squares) mark numerical results for $200$ eigenstates near $\overline{\epsilon}\approx \epsilon_{\rm max}/2$ ($\overline{\epsilon}\approx 0$).
Solid lines are results from Eq.~(\ref{PDF_2}), dashed lines are the rescaled results using Eq.~(\ref{eq:dist_rescaled2}).
}
\label{figdi2}
\end{figure}

Figures~\ref{figdi1} and~\ref{figdi2} compare the analytical expressions from Eqs.~(\ref{PDF_1})-(\ref{PDF_2}) to the actual numerical results.
At~$\Delta=0.1$, the agreement is reasonably good for both the diagonal matrix elements [Figs.~\ref{figdi1}(a) and~\ref{figdi1}(b)] and the offdiagonal matrix elements [Figs.~\ref{figdi2}(a) and~\ref{figdi2}(b)].
The deviations can be observed near the tails of the distributions.
In particular, the distributions calculated from the matrix elements between eigenstates in the middle of the spectrum ($\overline{\epsilon}\approx 0$) deviate less than those calculated from matrix elements between eigenstates away from the middle of the spectrum ($\overline{\epsilon}\approx \epsilon_{\rm max}/2$).

At $\Delta=0.5$, see Figs.~\ref{figdi1}(c)-\ref{figdi1}(d) and~\ref{figdi2}(c)-\ref{figdi2}(d), we only show the distributions away from the middle of the spectrum ($\overline{\epsilon}\approx \epsilon_{\rm max}/2$) since both diagonal and offdiagonal matrix elements are close to zero in the middle of the spectrum ($\overline{\epsilon}\approx 0$).
The predictions from Eqs.~(\ref{PDF_1})-(\ref{PDF_2}) are incompatible with the numerical results (see the solid lines).
This is a consequence of the variances being much smaller than the ones predicted by the simple RMT approximation.
For example, when the $\Delta=0.5$ case is compared to the $\Delta=0.1$ case, the ratio of their variances at $\overline{\epsilon}\approx \epsilon_{\rm max}/2$ for the largest system sizes $V\ge 18^3$ is given by
\begin{equation}
\varsigma^2_\text{diag/off}=\frac{\langle\langle\sigma^2_\text{diag/off}(\Delta=0.5)\rangle\rangle}{\langle\langle\sigma^2_\text{diag/off}(\Delta=0.1)\rangle\rangle}
\approx 10^{-1} \;.
\end{equation}
We then rescale the distributions according to the prescription
\begin{align} \label{eq:dist_rescaled1}
    & P_{\underline{n}_{\alpha\alpha}}\left(x\right)\rightarrow P_{\underline{n}_{\alpha\alpha}}\left(x/\varsigma_\text{diag}\right)/\varsigma_\text{diag}\;,\\
   & P_{\underline{n}_{\alpha\beta}}\left(x\right)\rightarrow P_{\underline{n}_{\alpha\beta}}\left(x/\varsigma_\text{off}\right)/\varsigma_\text{off}\;, \label{eq:dist_rescaled2}
\end{align}
which we plot as dashed lines in Figs.~\ref{figdi1}(c)-\ref{figdi1}(d) and~\ref{figdi2}(c)-\ref{figdi2}(d).
This rescaling improves the agreement with the numerical results.
However, the deviations near the tails of the distributions are still visible.
Nevertheless, they are similar to the deviations at $\Delta=0.1$ and $\overline{\epsilon}\approx \epsilon_{\rm max}/2$ shown in Figs.~\ref{figdi1}(a)-\ref{figdi1}(b) and~\ref{figdi2}(a)-\ref{figdi2}(b).

\subsection{Distributions of diagonal matrix elements} \label{sec:distribution_diag}

We first present the derivation of the distribution of $\underline{n}_{\alpha\alpha}$ from Eq.~(\ref{eq:def_naa_random}). Below, we show that is related to the distribution of a difference of random variables drawn from $\chi_1^2$, i.e., from the chi-square distribution of degree one.

The square of a random number from the Gaussian distribution belongs to the chi-square distribution for $x \ge 0$,
\begin{equation}
    P_{u^2}(x)=P_{v^2}(x)=\frac{1}{\sqrt{x}}\frac{1}{\sqrt{2\pi\sigma^2}}\exp(-x/{2\sigma^2})
\end{equation}
and vanishes for $x<0$ (see also Appendix~D of Ref.~\cite{lydzba_zhang_21}). Therefore, the distribution of a difference $w=u_{i\alpha}^2-v_{i\alpha}^2$ can be calculated as
\begin{equation}
\begin{split}
    P_w(y) & =\int_{-\infty}^\infty\int_{-\infty}^\infty dx\;dx^{'}\;
    P_{u^2}(x)P_{v^2}(x^{'})\delta(y-(x-x^{'}))\\
    & =\int_{-\infty}^\infty dx\;
    P_{u^2}(x)P_{v^2}(x-y),
\end{split}
\end{equation}
which takes the following form after introducing the chi-square distributions:
\begin{equation}
    P_w(y)=\frac{\exp(y/2\sigma^2)}{2\pi\sigma^2}\int_{\tilde{y}}^{\infty}dx\;\frac{\exp(-x/\sigma^2)}{\sqrt{x(x-y)}},
\end{equation}
where $\tilde{y}=y$ for $y\ge 0$ and $\tilde{y}=0$ for $y<0$. Finally, we arrive at
\begin{equation}
     P_w(y)=\frac{1}{2\pi\sigma^2}\text{K}_{0}(|y|/2\sigma^2),
\end{equation}
which is proportional to the Bessel function of the second kind $\text{K}_0$. Thus, the distribution of diagonal matrix elements $\underline{n}_{\alpha\alpha}$ is given by
\begin{equation}
    P_{\underline{n}_{\alpha\alpha}}(y)=\sqrt{N}P_{w}(\sqrt{N}y)=
    \frac{1}{\pi}\sqrt{\frac{V}{2}}\text{K}_0\left(\sqrt{\frac{V}{2}}|y|\right).
\end{equation}
Note that $P_{\underline{n}_{\alpha\alpha}}(x)$ is the same as the distribution of diagonal matrix elements of the nearest-neighbor hopping in the quantum-chaotic quadratic Hamiltonians that conserve the particle number~\cite{lydzba_zhang_21}. 

\subsection{Distributions of offdiagonal matrix elements} \label{sec:distribution_off}

Finally, we present the derivation of the distribution of $\underline{n}_{\alpha\beta}$ from Eq.~(\ref{eq:def_nab_random}). Below, we show that it is related to the distribution of a difference of two random variables from the function $\text{K}_0$, which itself is the distribution of a product of normal random variables.

The product of two random numbers from the Gaussian distribution is given by the Bessel function of the second kind $\text{K}_0$ (see also Appendix~D of Ref.~\cite{lydzba_zhang_21}),
\begin{equation}
    P_{uu}(x)=P_{vv}(x)=\frac{1}{\pi\sigma^2}\text{K}_0\left(|y|/\sigma^2\right).
\end{equation}
Therefore, the distribution of a difference $q=u_{i\alpha}u_{i\beta}-v_{i\alpha}v_{i\beta}$ can be calculated as
\begin{equation}
    P_q(y)=\int_{-\infty}^{\infty}dx\;P_{uu}(x)P_{vv}(x-y).
\end{equation}
This integral can be evaluated with the help of the so-called characteristic functions and the Fourier transform (for further details see Appendix~D of Ref.~\cite{lydzba_zhang_21}),
\begin{equation}
\begin{split}
    P_q(y) & =\frac{1}{\pi^2\sigma^4}\int_{-\infty}^{\infty}dx\;\text{K}_{0}(|x|/\sigma^2)\text{K}_{0}(|x-y|/\sigma^2)\\
    & = \frac{1}{2\sigma^2}\exp(-|y|/\sigma^2),
\end{split}
\end{equation}
and corresponds to the exponential distribution. Thus, the distribution of offdiagonal matrix elements $\underline{n}_{\alpha\beta}$ is given by
\begin{equation}
    P_{\underline{n}_{\alpha\beta}}(y)=\sqrt{N}P_q(\sqrt{N})=
    \sqrt{\frac{V}{2}}\exp(-\sqrt{2V}|y|).
\end{equation}
Note that $P_{\underline{n}_{\alpha\beta}}(x)$ is the same as the distribution of offdiagonal matrix elements of the nearest-neighbor hopping in the quantum-chaotic quadratic Hamiltonians that conserve the particle number~\cite{lydzba_zhang_21}.

\FloatBarrier
\bibliographystyle{biblev1}
\bibliography{references}

\end{document}